\documentclass[prd,aps,floatfix,twocolumn]{revtex4-1}
\usepackage{amsfonts,amsmath,amssymb,amsthm}
\usepackage{bm}
\usepackage{booktabs}
\usepackage{braket}
\usepackage{cases}
\usepackage{color}
\usepackage{dcolumn}
\usepackage{epsfig}
\usepackage{epstopdf}   
\usepackage{graphicx}
\usepackage[colorlinks,citecolor=blue,anchorcolor=red,menucolor=red,linkcolor=red,filecolor=red,runcolor=red,urlcolor=blue,frenchlinks=red]{hyperref}
\usepackage{indentfirst}
\usepackage{latexsym}
\usepackage{longtable}
\usepackage{multirow}
\usepackage{enumerate}
\usepackage{rotating}
\usepackage{slashed}  
\usepackage{subfigure}

\renewcommand{\arraystretch}{1.5}
\allowdisplaybreaks[4]
\newsavebox{\tablebox}
\usepackage{multirow}
\usepackage{dcolumn}
\usepackage{overpic}
\usepackage{booktabs}
\usepackage{makecell}
\usepackage{slashed}
\usepackage{epstopdf}
\usepackage{diagbox}
\usepackage{array}
\newcolumntype{|}{!{\vline}}
\begin{document}
\title{Where are the hidden-charm hexaquarks?}

\author{Zhe Liu$^{1,2}$}\email{zhliu20@lzu.edu.cn}
\author{Hong-Tao An$^{1,2}$}\email{anht14@lzu.edu.cn}
\author{Zhan-Wei Liu$^{1,2,3}$}\email{liuzhanwei@lzu.edu.cn}
\author{Xiang Liu$^{1,2,3}$}\email{xiangliu@lzu.edu.cn}
\affiliation{
$^1$School of Physical Science and Technology, Lanzhou University, Lanzhou 730000, China\\
$^2$Research Center for Hadron and CSR Physics, Lanzhou University and Institute of Modern Physics of CAS, Lanzhou 730000, China\\
$^3$Lanzhou Center for Theoretical Physics, Key Laboratory of Theoretical Physics of Gansu Province, and Frontiers Science Center for Rare Isotopes, Lanzhou University, Lanzhou 730000, China}

\date{\today}
\begin{abstract}
In this work, we carry out the study of hidden-charm hexaquark states with the typical configurations $qqc\bar{q}\bar{q}\bar{c}$ ($q=u, d, s$). The mass spectra of hidden-charm hexaquark states are obtained within the chromo-magnetic interaction model. In addition to the mass spectra analysis, we further illustrate their two-body strong decay behaviors. There exist some compact bound states which cannot decay through the strong interaction. Hopefully our results will help to search for such types of the exotic states in the future experiments.

\end{abstract}

\maketitle
\thispagestyle{empty} %

\section{Introduction}
\label{Sec:Introduction}
With the improvement of the luminosity and precision in experiment, more and more charmonium-like $XYZ$ states and $P_{c}$ states have been observed \cite{Choi:2003ue,Acosta:2003zx,Abazov:2004kp,Aaij:2014jqa,Ablikim:2016qzw,Ablikim:2017oaf,Ablikim:2020hsk,BESIII:2016adj,Aaij:2015tga,Aaij:2016phn,Aaij:2019vzc}. The present situation of hadronic states is far beyond the conventional quark model. The first doubly charm tetraquark $T_{\rm cc}^+$ with the configuration $cc\bar{u}\bar{d}$ was observed by the LHCb Collaboration \cite{Franz:2021talk}, and this newly discovered particle is explicitly an exotic state which cannot be classified into the conventional mesons.

The hexaquark states were proposed and the spectra of light-flavored hexaquarks were dynamically investigated very early after the birth of quark model. The $d^{*}(2380)$ resonance with $I(J^{P})=0(3^{+})$ has been reported by CELSIUS/WASA and WASA-at-COSY Collaborations \cite{Faldt:2011zv,Adlarson:2011bh,Adlarson:2012fe}, and it is expected to be a dibaryon which contains 6 constituent quarks. The deuteron is also a dibaryon. Jaffe firstly found the $H$ particle whose hyperfine interaction is much larger than that for two separated $\Lambda$ baryons within the chromo-magnetic interaction model \cite{Jaffe:1976yi}, and this dibaryon $uuddss$ was also studied within other framework \cite{Mackenzie:1985vv,Aerts:1984vv,Balachandran:1983dj,Straub:1988mz,Paganis:1999ux,Yost:1985mj,Rosner:1985yh,Karl:1987cg}. Moreover, the heavy dibaryons ($qqqqqQ$) \cite{Oka:2013iua,Gerasyuta:2011yg,Oka:2019mrd,Liu:2012zzo,Pepin:1998ih}, doubly-heavy dibaryons ($qqqqQQ$) \cite{Liu:2012zzo,Vijande:2016nzk,Wang:2017sto,Meng:2017fwb,Meguro:2011nr,Li:2012bt,Leandri:1995zm}, triply-heavy dibaryons ($qqqQQQ$) \cite{Wang:2020jqu,Chen:2018pzd,Richard:2020zxb}, the other fully light dibaryons ($qqqqqq$) \cite{Zhang:1997ny,Gerasyuta:2010hn,SilvestreBrac:1992yg,Park:2015nha,Oka:1988yq,Chen:2019vdh}, and even fully heavy dibaryons ($QQQQQQ$) \cite{Huang:2020bmb} were also proposed and discussed.

The hadronic states composed of three quarks and three antiquarks are another class of heaxquarks. The hidden-charm and hidden-bottom hexaquarks are especially focused on since they have much larger masses and thus are more easily distinguished from the ordinary mesons. With the hidden-charm tetraquark and pentaquark states observed in experiment, the discovery of hidden-charm hexaquarks would also come true in future.

Very recently, BESIII collaboration measured the cross section of the process $e^+e^- \rightarrow \pi^+ \pi^- \psi(3686)$ and further confirms the existence of three charmonium-like states wherein $Y(4660)$ is closed to the threshold of $\Lambda_{c}$-$\bar{\Lambda}_{c}$ systems \cite{BESIII:2021njb}. Before this, the structure $Y(4660)$ has been observed in the process of $e^+e^- \rightarrow \gamma_{\text{ISR}} \pi^+ \pi^- \psi(3686)$ in the Belle and BarBar experiments \cite{Belle:2007umv,Belle:2014wyt,BaBar:2012hpr}. $Y(4660)$ was interpreted as a higher charmonium in Ref. \cite{Wang:2020prx} and a hexaquark state configured by the triquark-antitriquark clusters in Ref. \cite{Qiao:2007ce}. The charmonium states can very likely be bound inside light hadronic matters, and such hadro-charmonium may explain the properties of the $Y(4660)$ peak \cite{Dubynskiy:2008mq}. G. Cotugno {\it et al.} suggested that the two observations of $Y(4660)$ and $Y(4630)$ are likely to be due to the same state constituted by four quarks in Ref. \cite{Cotugno:2009ys}.

The $\Lambda_{c}$-$\bar{\Lambda}_{c}$ structure was introduced to explain the production and decays of $Y(4260)$ in Refs. \cite{Qiao:2007ce,Qiao:2005av,Chen:2011cta}. $Y(4630)$ was observed in process $e^+e^- \rightarrow \Lambda_{c} \bar{\Lambda}_{c}$ in the Belle experiments \cite{Belle:2008xmh} and is considered as a candidate of $\Lambda_{c} \bar{\Lambda}_{c}$ bound state \cite{Lee:2011rka}. Especially, heavy baryon chiral perturbation theory was applied to systemically study the $\Lambda_{c}$-$\bar{\Lambda}_{c}$, $\Sigma_{c}$-$\bar{\Sigma}_{c}$, and $\Lambda_{b}$-$\bar{\Lambda}_{b}$ systems \cite{Chen:2013sba},
and the results suggest that $Y(4260)$ and $Y(4360)$ could be $\Lambda_{c}$-$\bar{\Lambda}_{c}$ baryonia. The two states are also suggested to be a mixture, with mixing close to maximal, of two states of hadrochamonium \cite{Li:2013ssa}.

The masses of baryonia with the open and hidden charm, bottomness and strangeness are studied in the framework of dispersion relation technique in Refs. \cite{Gerasyuta:2013esc,Gerasyuta:2020gyy,Gerasyuta:2020fii}.
The heavy baryon-antibaryon molecule states are investigated within the effective field theory \cite{Lu:2017dvm}. The hidden-charm and hidden-bottom hexaquark states were discussed within the QCD sum rules \cite{Wan:2019ake,Chen:2016ymy}. 

These work stimulate us to further study the hidden-charm hexaquark states. In this work we systemically investigate their mass spectra, stability, and two-body decay within the chromo-magnetic interaction (CMI) model.

The simple chromo-magnetic interaction arises from the one-gluon-exchange potential and further causes the mass splittings \cite{DeRujula:1975qlm,Liu:2019zoy}. The CMI model has been successfully adopted to study the mass spectra and stability of multiquark states \cite{Luo:2017eub,Wu:2016gas,Wu:2018xdi,Chen:2016ont,Wu:2016vtq,Liu:2016ogz,Wu:2017weo,Zhou:2018pcv,Li:2018vhp,An:2019idk,Cheng:2020irt,Cheng:2019obk,Hogaasen:2013nca,Weng:2018mmf,Weng:2019ynva,Weng:2020jao,Cheng:2020nho,An:2020jix,Karliner:2016zzc,Weng:2021hje,Zhao:2014qva}. The method can catch the basic features of hadron spectra, since the mass splittings between hadrons reflect the basic symmetries of their inner structures.

This paper is organized as follows. In Sec.~\ref{sec2}, the adopted CMI model and relevant parameters are introduced. We construct the flavor $\otimes$ color $\otimes$ spin wavefunctions for the $S$-wave hidden-charm hexaquark system in Sec. \ref{sec3}, and study the mass spectrum and the two-body decays through the strong interaction in Sec.~\ref{sec4}. A short summary follows in Sec. \ref{sec5}.

\section{THE Hamiltonian in the CMI model}
\label{sec2}
In the CMI model, the Hamiltonian has a simple form
\begin{eqnarray}\label{Eq1}
     H&=&\sum_i^6m_i+H_{\rm CMI},\nonumber\\
     H_{\rm CMI}&=&-\sum_{i<j}C_{ij} \bm{\lambda}_{i}\cdot\bm{\lambda}_{j}\bm{\sigma}_{i}\cdot\bm{\sigma}_{j},
\end{eqnarray}
where $m_i$ is the effective mass of the $i$-th constituent (anti) quark, and $\bm{\lambda}_{i}$ and $\bm{\sigma}_{i}$ are Gell-Mann and Pauli matrices, respectively. For the antiquark, $\bm{\lambda}_{\bar{q}}=-\bm{\lambda}_{q}^{*}$ and
$\bm{\sigma}_{\bar{q}}=\bm{\sigma}_{q}^{*}$. The dynamical effect of spatial wavefunctions plays an important role in the study of hadron spectrum. Chromomagnetic interaction is nonrelativistic in the Schr$\rm \ddot{o}$dinger equation in Ref. \cite{Godfrey:1985xj} wherein the authors used the spatial wave functions with harmonic-oscillator expansion. 
The $C_{ij}$ is effective coupling constant between the $i$-th (anti) quark and $j$-th (anti) quark
\begin{equation}
  C_{ij}= \frac{\pi \Braket{\alpha_s(r)\delta^3(\bm{r})}}{6m_i m_j},
\end{equation}
which is directly related to the spatial wavefunctions and the constituent quark masses. We focus on ground states in $S$-wave, and we simply suppose it does not change for various hexaquark systems.

H{\o}gaasen {\it et al.} found out that the $b$ quark mass in bottomonium is much lighter than the one in the heavy-light system, and introduced the color interaction (the spin-independent color Coulomb-like terms in the one-gluon-exchange interactions) in
Refs.~\cite{Hogaasen:2013nca,Karliner:2016zzc,Weng:2018mmf}.
We also introduce a color term into our model Refs.~\cite{Hogaasen:2013nca,Weng:2018mmf}
\begin{equation}
  H_{\text{C}} = -\sum_{i<j} A_{ij}
  \bm\lambda_i \cdot \bm\lambda_j.
\end{equation}
The nonvanishing color interaction coefficient $A_{ij}$ implies a change of the effective masses. We can rewrite the CMI Hamiltonian as Ref.~\cite{Weng:2018mmf}
\begin{eqnarray}
H = -\frac{3}{4}\sum_{i<j} m_{ij} \bm{\lambda}_{i} \cdot \bm{\lambda}_{j}- \sum_{i<j} v_{ij} \bm{\lambda}_{i} \cdot \bm{\lambda}_{j} \bm{\sigma}_{i} \cdot \bm{\sigma}_{j} ,
\end{eqnarray}
where
\begin{eqnarray}
  m_{ij} = \frac{1}{4}\left( m_i + m_j\right) + \frac{4}3 A_{ij}.
\end{eqnarray}

To estimate the mass spectra of the hidden-charm hexaquark states, we extract the effective coupling parameters $m_{ij}$ and $v_{ij}$ from the conventional hadron masses \cite{Weng:2018mmf}. In the present work, $v_{q\bar q}$ and $m_{q\bar q}$ are only determined by vector mesons ($q=n,s$ and $n=u,d$). We present the obtained effective coupling parameters in Table \ref{parameter}.

\begin{table}
	\centering
	\caption{The effective coupling parameters in units of MeV.}
	\label{parameter}
	\begin{tabular}{cccc|ccccc}
\bottomrule[1.5pt]
\bottomrule[0.5pt]
$m_{nn}$&$m_{ns}$&$m_{ss}$&$m_{nc}$&$m_{n\bar{n}}$&$m_{n\bar{s}}$&$m_{s\bar{s}}$&$m_{n\bar{c}}$&$m_{c\bar{c}}$\\
182.2&226.7&262.3&520.0&166.49&204.2&241.1&493.3&767.1\\
\bottomrule[0.7pt]
$v_{nn}$&$v_{ns}$&$v_{ss}$&$v_{nc}$&$v_{n\bar{n}}$&$v_{n\bar{s}}$&$v_{s\bar{s}}$&$v_{n\bar{c}}$&$v_{c\bar{c}}$\\
19.1&13.3&12.2&3.9&20.5&14.2&10.3&6.6&5.3\\
\bottomrule[0.5pt]
\bottomrule[1.5pt]
	\end{tabular}
\end{table}

 \section{The wavefunctions}
\label{sec3}
In order to calculate the CMI Hamiltonian, we need to exhaust all the possible spin and color wavefunctions of hexaquark states and combine them with the corresponding flavor wavefunctions. The constructed flavor-color-spin wavefunctions should be fully antisymmetric when exchanging identical quarks because of Pauli principle. The wavefunctions do not change with different sets of basis, and we use the $|[(q_{1}q_{2})c] [(\bar{q}_{3}\bar{q}_{4})\bar{c}]\rangle$ basis to construct the hidden-charm hexaquarks wavefunctions.

Firstly, we discuss the flavor wavefunctions. The mass hierarchy for $c$, $s$ and $ud$ quarks is obvious and we neglect the mixing effect among the $c\bar c$, $s\bar s$, and $n\bar n$ pairs. Based on these, we list all the possible flavor combinations for the hidden-charm hexaquark system in Table \ref{flavor}.

In Table \ref{flavor}, the three subsystems of the first line are pure neutral particles and $C$ parity is ``good" quantum number. For the six subsystems of the second line, every subsystem has a charge conjugation anti-partner, thus they have the same mass spectra, and we only need to discuss one of two relevant subsystems. In the first line of Table \ref{flavor}, $nnc\bar{n}\bar{n}\bar{c}$ has isospin $I=(2, 1, 0)$ and $nsc\bar{n}\bar{s}\bar{c}$ has isospin $I=(1, 0)$. In the second line, the isospin $I$ can be $(3/2, 1/2)$ for $nsc\bar{n}\bar{n}\bar{c}$, $(1, 0)$ for $nnc\bar{s}\bar{s}\bar{c}$, and 1/2 for $nsc\bar{s}\bar{s}\bar{c}$.

\begin{table}[t]
\centering \caption{All possible flavor combinations for the hidden-charm hexaquark system.
}\label{flavor}
\renewcommand\arraystretch{1.4}
\begin{tabular}{c|ccccc}
\bottomrule[1.5pt]
\bottomrule[0.5pt]
System&\multicolumn{5}{c}{Flavor combinations}\\
\bottomrule[0.5pt]
\multirow{2}*{$qqc\bar{q}\bar{q}\bar{c}$}&$nnc\bar{n}\bar{n}\bar{c}$&\quad&$ssc\bar{s}\bar{s}\bar{c}$&\quad&$nsc\bar{n}\bar{s}\bar{c}$\\
&$nsc\bar{n}\bar{n}\bar{c}$ ($nnc\bar{n}\bar{s}\bar{c}$)&\quad&$nnc\bar{s}\bar{s}\bar{c}$  ($ssc\bar{n}\bar{n}\bar{c}$)&\quad&$nsc\bar{s}\bar{s}\bar{c}$  ($ssc\bar{n}\bar{s}\bar{c}$)\\
\bottomrule[0.5pt]
\bottomrule[1.5pt]
\end{tabular}
\end{table}

Next, we briefly introduce the color wavefunctions for all hexaquark systems.
They can be deduced from the following direct product:

\begin{equation}
\scalebox{0.863}{
\mbox{
$
\begin{split}
&([3]\otimes[3]\otimes[3])\otimes([\bar{3}]\otimes[\bar{3}]\otimes[\bar{3}])\\
=&([1_{\rm A}]\oplus[8_{\rm MA}]\oplus[8_{\rm MS}]\oplus[10_{\rm S}])\otimes([1_{\rm A}]\oplus[8_{\rm MA}]\oplus[8_{\rm MS}]\oplus[\bar{10}_{S}])\\
\rightarrow&([1_{\rm A}]\otimes[1_{\rm A}])\oplus([8_{\rm MA}]\otimes[8_{\rm MA}])\oplus([8_{\rm MS}]\otimes[8_{\rm MA}])\oplus\\
&([8_{\rm MA}]\otimes[8_{\rm MS}])\oplus([8_{\rm MS}]\otimes[8_{\rm MS}])\oplus([10_{\rm S}]\otimes[\bar{10}_{\rm S}]),
\end{split}
$
}}
\end{equation}
where A (S) means totally symmetric (antisymmetric), and MS (MA) means that $q_{1}q_{2}$ or $\bar{q}_{3}\bar{q}_{4}$ is symmetric (antisymmetric).
Here, the color-singlet wavefunctions for the hexaquarks are shown in Table \ref{color}.
In the notation $|[(q_{1}q_{2})^{\rm color1}c ]^{\rm color3} [ (\bar{q}_{3}\bar{q}_{4})^{\rm color2}\bar{c}]^{\rm color4}\rangle$, the color1, color2, color3, and color4 stand for the color representations of $q_{1}q_{2}$, $\bar{q}_{3}\bar{q}_{4}$, $q_{1}q_{2}c$, and $\bar{q}_{3}\bar{q}_{4}\bar{c}$, respectively.

\begin{table}[t]
\centering \caption{All possible color and spin wavefunctions for the hidden-charm hexaquark system.
}\label{color}
\begin{lrbox}{\tablebox}
\renewcommand\arraystretch{1.4}
\begin{tabular}{ll}
\bottomrule[1.5pt]
\bottomrule[0.5pt]
Color wavefunctions\\
\bottomrule[0.5pt]
$\phi_{1}^{\rm AA} =  |[(q_{1}q_{2})^{\bar{3}}c ]^{1} [ (\bar{q}_{3}\bar{q}_{4})^{3}\bar{c}]^{1}\rangle$ &
$\phi_{2}^{\rm MAMA} =  |[(q_{1}q_{2})^{\bar{3}}c ]^{8} [ (\bar{q}_{3}\bar{q}_{4})^{3}\bar{c} ]^{8}\rangle$\\
$\phi_{3}^{\rm MSMA} =  |[(q_{1}q_{2})^{6}c ]^{8} [ (\bar{q}_{3}\bar{q}_{4})^{3}\bar{c} ]^{8}\rangle$&
$\phi_{4}^{\rm MAMS} =  |[(q_{1}q_{2})^{\bar{3}}c ]^{8} [ (\bar{q}_{3}\bar{q}_{4})^{\bar{6}}\bar{c} ]^{8}\rangle$\\
$\phi_{5}^{\rm MSMS} =  |[(q_{1}q_{2})^{6}c ]^{8} [ (\bar{q}_{3}\bar{q}_{4})^{\bar{6}}\bar{c} ]^{8}\rangle$&
$\phi_{6}^{\rm SS} =  |[(q_{1}q_{2})^{6}c ]^{10} [ (\bar{q}_{3}\bar{q}_{4})^{\bar{6}}\bar{c} ]^{\bar{10}}\rangle$\\	
\bottomrule[1.0pt]
Spin wavefunctions\\
\bottomrule[0.5pt]
Spin=0:\\
$\chi_{1}^{\rm MSMS} =  |[(q_{1}q_{2})_{1}c ]_{\frac{1}{2}} [ (\bar{q}_{3}\bar{q}_{4})_{1}\bar{c} ]_{\frac{1}{2}}\rangle_{0}$&
$\chi_{2}^{\rm SS} =  |[(q_{1}q_{2})_{1}c ]_{\frac{3}{2}} [ (\bar{q}_{3}\bar{q}_{4})_{1}\bar{c} ]_{\frac{3}{2}}\rangle_{0}$
\\
$\chi_{3}^{\rm MSA} =  |[(q_{1}q_{2})_{1}c ]_{\frac{1}{2}} [ (\bar{q}_{3}\bar{q}_{4})_{0}\bar{c} ]_{\frac{1}{2}}\rangle_{0}$&
$\chi_{4}^{\rm A MS} =  |[(q_{1}q_{2})_{0}c ]_{\frac{1}{2}} [ (\bar{q}_{3}\bar{q}_{4})_{1}\bar{c} ]_{\frac{1}{2}}\rangle_{0}$
\\	
$\chi_{5}^{\rm AA} =  |[(q_{1}q_{2})_{0}c ]_{\frac{1}{2}} [ (\bar{q}_{3}\bar{q}_{4})_{0}\bar{c} ]_{\frac{1}{2}}\rangle_{0}$\\
\bottomrule[0.5pt]
Spin=1:\\
$\chi_{6}^{\rm MSMS} =  |[(q_{1}q_{2})_{1}c ]_{\frac{1}{2}} [ (\bar{q}_{3}\bar{q}_{4})_{1}\bar{c} ]_{\frac{1}{2}}\rangle_{1}$&
$\chi_{7}^{\rm SS} =  |[(q_{1}q_{2})_{1}c ]_{\frac{3}{2}} [ (\bar{q}_{3}\bar{q}_{4})_{1}\bar{c} ]_{\frac{3}{2}}\rangle_{1}$
\\
$\chi_{8}^{\rm MSS} =  |[(q_{1}q_{2})_{1}c ]_{\frac{3}{2}} [ (\bar{q}_{3}\bar{q}_{4})_{1}\bar{c} ]_{\frac{1}{2}}\rangle_{1}$&
$\chi_{9}^{\rm SMS} =  |[(q_{1}q_{2})_{1}c ]_{\frac{1}{2}} [ (\bar{q}_{3}\bar{q}_{4})_{1}\bar{c} ]_{\frac{3}{2}}\rangle_{1}$
\\
$\chi_{10}^{\rm MSA} =  |[(q_{1}q_{2})_{1}c ]_{\frac{1}{2}} [ (\bar{q}_{3}\bar{q}_{4})_{0}\bar{c} ]_{\frac{1}{2}}\rangle_{1}$&
$\chi_{11}^{\rm SA} =  |[(q_{1}q_{2})_{1}c ]_{\frac{3}{2}} [ (\bar{q}_{3}\bar{q}_{4})_{0}\bar{c} ]_{\frac{1}{2}}\rangle_{1}$
\\
$\chi_{12}^{\rm A MS} =  |[(q_{1}q_{2})_{0}c ]_{\frac{1}{2}} [ (\bar{q}_{3}\bar{q}_{4})_{1}\bar{c} ]_{\frac{1}{2}}\rangle_{1}$&
$\chi_{13}^{\rm AS} =  |[(q_{1}q_{2})_{0}c ]_{\frac{1}{2}} [ (\bar{q}_{3}\bar{q}_{4})_{1}\bar{c} ]_{\frac{3}{2}}\rangle_{1}$
\\	
$\chi_{14}^{\rm AA} =  |[(q_{1}q_{2})_{0}c ]_{\frac{1}{2}} [ (\bar{q}_{3}\bar{q}_{4})_{0}\bar{c} ]_{\frac{1}{2}}\rangle_{1}$\\
\bottomrule[0.5pt]
Spin=2:\\
$\chi_{15}^{\rm SS} =  |[(q_{1}q_{2})_{1}c ]_{\frac{3}{2}} [ (\bar{q}_{3}\bar{q}_{4})_{1}\bar{c} ]_{\frac{3}{2}}\rangle_{2}$&
$\chi_{16}^{\rm SMS} =  |[(q_{1}q_{2})_{1}c ]_{\frac{3}{2}} [ (\bar{q}_{3}\bar{q}_{4})_{1}\bar{c} ]_{\frac{1}{2}}\rangle_{2}$
\\
$\chi_{17}^{\rm MSS} =  |[(q_{1}q_{2})_{1}c ]_{\frac{1}{2}} [ (\bar{q}_{3}\bar{q}_{4})_{1}\bar{c} ]_{\frac{3}{2}}\rangle_{2}$&
$\chi_{18}^{\rm SA} =  |[(q_{1}q_{2})_{1}c ]_{\frac{3}{2}} [ (\bar{q}_{3}\bar{q}_{4})_{0}\bar{c} ]_{\frac{1}{2}}\rangle_{2}$
\\
$\chi_{19}^{\rm AS} =  |[(q_{1}q_{2})_{0}c ]_{\frac{1}{2}} [ (\bar{q}_{3}\bar{q}_{4})_{1}\bar{c} ]_{\frac{3}{2}}\rangle_{2}$
\\
\bottomrule[0.5pt]
Spin=3:\\
$\chi_{20}^{\rm SS} =  |[(q_{1}q_{2})_{1}c ]_{\frac{3}{2}} [ (\bar{q}_{3}\bar{q}_{4})_{1}\bar{c} ]_{\frac{3}{2}}\rangle_{3}$\\
\bottomrule[0.5pt]
\bottomrule[1.5pt]
\end{tabular}
\end{lrbox}\scalebox{0.90}{\usebox{\tablebox}}
\end{table}

Lastly, the spin wavefunctions for the hidden-charm hexaquark states are also shown in Table \ref{color}.
In the notation $|[(q_{1}q_{2})_{\rm spin1}c ]_{\rm spin3} [ (\bar{q}_{3}\bar{q}_{4})_{\rm spin2}\bar{c}]_{\rm spin4}\rangle_{\rm spin5}$, the spin1, spin2, spin3, spin4, and spin5 represent the spins of $q_{1}q_{2}$, $\bar{q}_{3}\bar{q}_{4}$, $q_{1}q_{2}c$, $\bar{q}_{3}\bar{q}_{4}\bar{c}$, and the total spin, respectively.

Considering the Pauli principle, we obtain 54 types of total wavefunctions and present them in the first part of Table \ref{type}. Some wavefunctions are the eigenstates of $C$ parity like $[\phi^{\rm SS}\otimes\chi^{\rm SS}]$, but others are not. For the neutral states,  we need do linear superposition to construct eigen wavefunctions of $C$ parity, and present them in the second part of Table \ref{type}. We introduce notations $\delta_{12}^{A}$, $\delta_{12}^{S}$, $\delta_{34}^{A}$, and $\delta_{34}^{S}$. When the two light quarks or antiquarks are antisymmetric (symmetric) in the flavor space, $\delta_{12}^{A}=0$ ($\delta_{12}^{S}=0$), or else $\delta_{12}^{A}=1$ ($\delta_{12}^{S}=1$).  The hidden-charm hexaquark states can be categorized into 6 classes, and we present them in third part of Table \ref{type}.

\begin{table*}[t]
\centering \caption{All possible types of total wavefunctions and different classes of the hidden-charm hexaquark system
}\label{type}
\begin{lrbox}{\tablebox}
\renewcommand\arraystretch{1.4}
\begin{tabular}{llllll}
\bottomrule[1.5pt]
\bottomrule[0.5pt]
\multicolumn{6}{l}{All possible types of total wavefunctions for hexaquark system without $C$ parity}\\
\bottomrule[0.5pt]
$[\phi^{\rm AA}\otimes\chi^{\rm SS}]\delta_{12}^{\rm A}\delta_{34}^{\rm A}$&$[\phi^{\rm MAMA}\otimes\chi^{\rm SS}]\delta_{12}^{\rm A}\delta_{34}^{\rm A}$&$[\phi^{\rm MSMA}\otimes\chi^{\rm SS}]\delta_{12}^{\rm S}\delta_{34}^{\rm A}$&
$[\phi^{\rm SS}\otimes\chi^{\rm SS}]\delta_{12}^{\rm S}\delta_{34}^{\rm S}$&$[\phi^{\rm MSMA}\otimes\chi^{\rm SS}]\delta_{12}^{\rm S}\delta_{34}^{\rm A}$&$[\phi^{\rm MSMS}\otimes\chi^{\rm SS}]\delta_{12}^{\rm S}\delta_{34}^{\rm S}$\\
$[\phi^{\rm AA}\otimes\chi^{\rm SA}]\delta_{12}^{\rm A}\delta_{34}^{\rm S}$&$[\phi^{\rm MAMA}\otimes\chi^{\rm SA}]\delta_{12}^{\rm A}\delta_{34}^{\rm S}$&$[\phi^{\rm MSMA}\otimes\chi^{\rm SA}]\delta_{12}^{\rm S}\delta_{34}^{\rm S}$&
$[\phi^{\rm SS}\otimes\chi^{\rm SA}]\delta_{12}^{\rm S}\delta_{34}^{\rm A}$&$[\phi^{\rm MSMA}\otimes\chi^{\rm SA}]\delta_{12}^{\rm S}\delta_{34}^{\rm S}$&$[\phi^{\rm MSMS}\otimes\chi^{\rm SA}]\delta_{12}^{\rm S}\delta_{34}^{\rm A}$\\
$[\phi^{\rm AA}\otimes\chi^{\rm AS}]\delta_{12}^{\rm S}\delta_{34}^{\rm A}$&$[\phi^{\rm MAMA}\otimes\chi^{\rm AS}]\delta_{12}^{\rm S}\delta_{34}^{\rm A}$&$[\phi^{\rm MSMA}\otimes\chi^{\rm AS}]\delta_{12}^{\rm A}\delta_{34}^{\rm A}$&
$[\phi^{\rm SS}\otimes\chi^{\rm AS}]\delta_{12}^{\rm A}\delta_{34}^{\rm S}$&$[\phi^{\rm MSMA}\otimes\chi^{\rm AS}]\delta_{12}^{\rm A}\delta_{34}^{\rm A}$&$[\phi^{\rm MSMS}\otimes\chi^{\rm AS}]\delta_{12}^{\rm A}\delta_{34}^{\rm S}$\\
$[\phi^{\rm AA}\otimes\chi^{\rm AA}]\delta_{12}^{\rm S}\delta_{34}^{\rm S}$&$[\phi^{\rm MAMA}\otimes\chi^{\rm AA}]\delta_{12}^{\rm S}\delta_{34}^{\rm S}$&$[\phi^{\rm MSMA}\otimes\chi^{\rm AA}]\delta_{12}^{\rm A}\delta_{34}^{\rm S}$&
$[\phi^{\rm SS}\otimes\chi^{\rm AA}]\delta_{12}^{\rm A}\delta_{34}^{\rm A}$&$[\phi^{\rm MSMA}\otimes\chi^{\rm AA}]\delta_{12}^{\rm A}\delta_{34}^{\rm S}$&$[\phi^{\rm MSMS}\otimes\chi^{\rm AA}]\delta_{12}^{\rm A}\delta_{34}^{\rm A}$\\
$[\phi^{\rm AA}\otimes\chi^{\rm SMS}]\delta_{12}^{\rm A}\delta_{34}^{\rm A}$&$[\phi^{\rm MAMA}\otimes\chi^{\rm SMS}]\delta_{12}^{\rm A}\delta_{34}^{\rm A}$&$[\phi^{\rm MSMA}\otimes\chi^{\rm SMS}]\delta_{12}^{\rm S}\delta_{34}^{\rm A}$&
$[\phi^{\rm SS}\otimes\chi^{\rm SMS}]\delta_{12}^{\rm S}\delta_{34}^{\rm S}$&$[\phi^{\rm MSMA}\otimes\chi^{\rm SMS}]\delta_{12}^{\rm S}\delta_{34}^{\rm A}$&$[\phi^{\rm MSMS}\otimes\chi^{\rm SMS}]\delta_{12}^{\rm S}\delta_{34}^{\rm S}$\\
$[\phi^{\rm AA}\otimes\chi^{\rm MSS}]\delta_{12}^{\rm A}\delta_{34}^{\rm A}$&$[\phi^{\rm MAMA}\otimes\chi^{\rm MSS}]\delta_{12}^{\rm A}\delta_{34}^{\rm A}$&$[\phi^{\rm MSMA}\otimes\chi^{\rm MSS}]\delta_{12}^{\rm S}\delta_{34}^{\rm A}$&
$[\phi^{\rm SS}\otimes\chi^{\rm MSS}]\delta_{12}^{\rm S}\delta_{34}^{\rm S}$&$[\phi^{\rm MSMA}\otimes\chi^{\rm MSS}]\delta_{12}^{\rm S}\delta_{34}^{\rm A}$&$[\phi^{\rm MSMS}\otimes\chi^{\rm MSS}]\delta_{12}^{\rm S}\delta_{34}^{\rm S}$\\
$[\phi^{\rm AA}\otimes\chi^{\rm MSA}]\delta_{12}^{\rm A}\delta_{34}^{\rm S}$&$[\phi^{\rm MAMA}\otimes\chi^{\rm MSA}]\delta_{12}^{\rm A}\delta_{34}^{\rm S}$&$[\phi^{\rm MSMA}\otimes\chi^{\rm MSA}]\delta_{12}^{\rm S}\delta_{34}^{\rm S}$&
$[\phi^{\rm SS}\otimes\chi^{\rm MSA}]\delta_{12}^{\rm S}\delta_{34}^{\rm A}$&$[\phi^{\rm MSMA}\otimes\chi^{\rm MSA}]\delta_{12}^{\rm S}\delta_{34}^{\rm S}$&$[\phi^{\rm MSMS}\otimes\chi^{\rm MSA}]\delta_{12}^{\rm S}\delta_{34}^{\rm A}$\\
$[\phi^{\rm AA}\otimes\chi^{\rm A MS}]\delta_{12}^{\rm S}\delta_{34}^{\rm A}$&$[\phi^{\rm MAMA}\otimes\chi^{\rm A MS}]\delta_{12}^{\rm S}\delta_{34}^{\rm A}$&$[\phi^{\rm MSMA}\otimes\chi^{\rm A MS}]\delta_{12}^{\rm A}\delta_{34}^{\rm A}$&
$[\phi^{\rm SS}\otimes\chi^{\rm A MS}]\delta_{12}^{\rm A}\delta_{34}^{\rm S}$&$[\phi^{\rm MSMA}\otimes\chi^{\rm A MS}]\delta_{12}^{\rm A}\delta_{34}^{\rm A}$&$[\phi^{\rm MSMS}\otimes\chi^{\rm A MS}]\delta_{12}^{\rm A}\delta_{34}^{\rm S}$\\
$[\phi^{\rm AA}\otimes\chi^{\rm MSMS}]\delta_{12}^{\rm A}\delta_{34}^{\rm A}$&$[\phi^{\rm MAMA}\otimes\chi^{\rm MSMS}]\delta_{12}^{\rm A}\delta_{34}^{\rm A}$&$[\phi^{\rm MSMA}\otimes\chi^{\rm MSMS}]\delta_{12}^{\rm S}\delta_{34}^{\rm A}$&
$[\phi^{\rm SS}\otimes\chi^{\rm MSMS}]\delta_{12}^{\rm S}\delta_{34}^{\rm S}$&$[\phi^{\rm MSMA}\otimes\chi^{\rm MSMS}]\delta_{12}^{\rm S}\delta_{34}^{\rm A}$&$[\phi^{\rm MSMS}\otimes\chi^{\rm MSMS}]\delta_{12}^{\rm S}\delta_{34}^{\rm S}$\\
\bottomrule[1.0pt]
\multicolumn{6}{l}{All possible types of total wavefunctions for pure neutral hexaquark system}\\
\bottomrule[0.5pt]
$[\phi^{\rm SS}\otimes\chi^{\rm SS}]\delta_{12}^{\rm S}\delta_{34}^{\rm S}$&
$[\phi^{\rm SS}\otimes\chi^{\rm MSMS}]\delta_{12}^{\rm S}\delta_{34}^{\rm S}$&
$[\phi^{\rm SS}\otimes\chi^{\rm AA}]\delta_{12}^{\rm A}\delta_{34}^{\rm A}$&
$[\phi^{\rm MSMS}\otimes\chi^{\rm SS}]\delta_{12}^{\rm S}\delta_{34}^{\rm S}$&
$[\phi^{\rm MSMS}\otimes\chi^{\rm AA}]\delta_{12}^{\rm A}\delta_{34}^{\rm A}$&
$[\phi^{\rm MSMS}\otimes\chi^{\rm MSMS}]\delta_{12}^{\rm S}\delta_{34}^{\rm S}$
\\
$[\phi^{\rm AA}\otimes\chi^{\rm SS}]\delta_{12}^{\rm A}\delta_{34}^{\rm A}$&
$[\phi^{\rm AA}\otimes\chi^{\rm MSMS}]\delta_{12}^{\rm A}\delta_{34}^{\rm A}$&
$[\phi^{\rm AA}\otimes\chi^{\rm AA}]\delta_{12}^{\rm S}\delta_{34}^{\rm S}$&
$[\phi^{\rm MAMA}\otimes\chi^{\rm SS}]\delta_{12}^{\rm A}\delta_{34}^{\rm A}$&
$[\phi^{\rm MAMA}\otimes\chi^{\rm AA}]\delta_{12}^{\rm S}\delta_{34}^{\rm S}$&
$[\phi^{\rm MAMA}\otimes\chi^{\rm MSMS}]\delta_{12}^{\rm A}\delta_{34}^{\rm A}$
\\
\multicolumn{2}{l}{$\frac{1}{\sqrt{2}}[(\phi^{\rm MAMS}\pm\phi^{\rm MSMA})\otimes\chi^{\rm SS}]\delta_{12}^{\rm S}\delta_{12}^{\rm A}\delta_{34}^{\rm S}\delta_{34}^{\rm A}$}
&\multicolumn{2}{l}{$\frac{1}{\sqrt{2}}[(\phi^{\rm MAMS}\pm\phi^{\rm MSMA})\otimes\chi^{\rm MSMS}]\delta_{12}^{\rm S}\delta_{12}^{\rm A}\delta_{34}^{\rm S}\delta_{34}^{\rm A}$}
&\multicolumn{2}{l}{$\frac{1}{\sqrt{2}}[(\phi^{\rm MAMS}\pm\phi^{\rm MSMA})\otimes\chi^{\rm AA}]\delta_{12}^{\rm S}\delta_{12}^{\rm A}\delta_{34}^{\rm S}\delta_{34}^{\rm A}$}
\\
\multicolumn{2}{l}{$\frac{1}{\sqrt{2}}[\phi^{\rm SS}\otimes(\chi^{\rm SA}\pm\chi^{\rm AS})]\delta_{12}^{\rm S}\delta_{12}^{\rm A}\delta_{34}^{\rm S}\delta_{34}^{\rm A}$}
&\multicolumn{2}{l}{$\frac{1}{\sqrt{2}}[\phi^{\rm AA}\otimes(\chi^{\rm SA}\pm\chi^{\rm AS})]\delta_{12}^{\rm S}\delta_{12}^{\rm A}\delta_{34}^{\rm S}\delta_{34}^{\rm A}$}
&\multicolumn{2}{l}{$\frac{1}{\sqrt{2}}[\phi^{\rm MSMS}\otimes(\chi^{\rm SA}\pm\chi^{\rm AS})]\delta_{12}^{\rm S}\delta_{12}^{\rm A}\delta_{34}^{\rm S}\delta_{34}^{\rm A}$}
\\
\multicolumn{2}{l}{$\frac{1}{\sqrt{2}}[\phi^{\rm SS}\otimes(\chi^{\rm MSA}\pm\chi^{\rm A MS})]\delta_{12}^{\rm S}\delta_{12}^{\rm A}\delta_{34}^{\rm S}\delta_{34}^{\rm A}$}
&\multicolumn{2}{l}{$\frac{1}{\sqrt{2}}[\phi^{\rm AA}\otimes(\chi^{\rm MSA}\pm\chi^{\rm A MS})]\delta_{12}^{\rm S}\delta_{12}^{\rm A}\delta_{34}^{\rm S}\delta_{34}^{\rm A}$}
&\multicolumn{2}{l}{$\frac{1}{\sqrt{2}}[\phi^{\rm MSMS}\otimes(\chi^{\rm MSA}\pm\chi^{\rm A MS})]\delta_{12}^{\rm S}\delta_{12}^{\rm A}\delta_{34}^{\rm S}\delta_{34}^{\rm A}$}
\\
\multicolumn{2}{l}{$\frac{1}{\sqrt{2}}[\phi^{\rm SS}\otimes(\chi^{\rm SMS}\pm\chi^{\rm MSS})]\delta_{12}^{\rm S}\delta_{34}^{\rm S}$}
&\multicolumn{2}{l}{$\frac{1}{\sqrt{2}}[\phi^{\rm AA}\otimes(\chi^{\rm SMS}\pm\chi^{\rm SMS})]\delta_{12}^{\rm A}\delta_{34}^{\rm A}$}
&\multicolumn{2}{l}{$\frac{1}{\sqrt{2}}[\phi^{\rm MSMS}\otimes(\chi^{\rm SMS}\pm\chi^{\rm SMS})]\delta_{12}^{\rm S}\delta_{34}^{\rm S}$}
\\
\multicolumn{2}{l}{$\frac{1}{\sqrt{2}}[\phi^{\rm MAMA}\otimes(\chi^{\rm MSA}\pm\chi^{\rm A MS})]\delta_{12}^{\rm S}\delta_{12}^{\rm A}\delta_{34}^{\rm S}\delta_{34}^{\rm A}$}
&\multicolumn{2}{l}{$\frac{1}{\sqrt{2}}[\phi^{\rm MAMA}\otimes(\chi^{\rm SA}\pm\chi^{\rm AS})]\delta_{12}^{\rm S}\delta_{12}^{\rm A}\delta_{34}^{\rm S}\delta_{34}^{\rm A}$}
&\multicolumn{2}{l}{$\frac{1}{\sqrt{2}}[\phi^{\rm MAMA}\otimes(\chi^{\rm SMS}\pm\chi^{\rm SMS})]\delta_{12}^{\rm A}\delta_{34}^{\rm A}$}
\\
\multicolumn{3}{l}{$\frac{1}{\sqrt{2}}[(\phi^{\rm MAMS}\otimes\chi^{\rm SA})\pm(\phi^{\rm MSMA}\otimes\chi^{\rm AS})]\delta_{12}^{\rm S}\delta_{12}^{\rm A}\delta_{34}^{\rm S}\delta_{34}^{\rm A}$}
&\multicolumn{3}{l}{$\frac{1}{\sqrt{2}}[(\phi^{\rm MAMS}\otimes\chi^{\rm MSA})\pm(\phi^{\rm MSMA}\otimes\chi^{\rm A MS})]\delta_{12}^{\rm S}\delta_{12}^{\rm A}\delta_{34}^{\rm S}\delta_{34}^{\rm A}$}
\\
\multicolumn{3}{l}{$\frac{1}{\sqrt{2}}[(\phi^{\rm MAMS}\otimes\chi^{\rm AS})\pm(\phi^{\rm MSMA}\otimes\chi^{\rm SA})]\delta_{12}^{\rm S}\delta_{12}^{\rm A}\delta_{34}^{\rm S}\delta_{34}^{\rm A}$}
&\multicolumn{3}{l}{$\frac{1}{\sqrt{2}}[(\phi^{\rm MAMS}\otimes\chi^{\rm MSA})\pm(\phi^{\rm MSMA}\otimes\chi^{\rm A MS})]\delta_{12}^{\rm S}\delta_{12}^{\rm A}\delta_{34}^{\rm S}\delta_{34}^{\rm A}$}
\\
\bottomrule[1.0pt]
\multicolumn{6}{l}{Different classes of the hidden-charm hexaquark system}\\
\bottomrule[0.5pt]
\multicolumn{2}{c}{\multirow{2}*{\quad$\delta_{12}^{\rm A}=1$, $\delta_{34}^{\rm A}=1$, $\delta_{12}^{\rm S}=0$, $\delta_{34}^{\rm S}=0$ :}}& \multicolumn{1}{l|}{$(nn)^{ I=1}c(\bar{n}\bar{n})^{ I=1}\bar{c}$,}&
\multicolumn{2}{c}{\multirow{2}*{\quad$\delta_{12}^{\rm A}=0$, $\delta_{34}^{\rm A}=1$, $\delta_{12}^{\rm S}=1$, $\delta_{34}^{\rm S}=0$ :}}& \multicolumn{1}{l}{$(nn)^{I=0}c(\bar{n}\bar{n})^{I=1}\bar{c}$,}\\
&&\multicolumn{1}{l|}{$(nn)^{ I=1}c\bar{s}\bar{s}\bar{c}$,\quad $ssc\bar{s}\bar{s}\bar{c}$}&&&\multicolumn{1}{l}{$(nn)^{I=0}c\bar{s}\bar{s}\bar{c}$}\\
\bottomrule[0.1pt]
\multicolumn{2}{c}{\quad$\delta_{12}^{\rm A}=0$, $\delta_{34}^{\rm A}=0$, $\delta_{12}^{\rm S}=1$, $\delta_{34}^{\rm S}=1$ :}&
\multicolumn{1}{l|}{$(nn)^{I=0}c(\bar{n}\bar{n})^{ I=0}\bar{c}$}&
\multicolumn{2}{c}{\quad$\delta_{12}^{\rm A}=1$, $\delta_{34}^{\rm A}=1$, $\delta_{12}^{\rm S}=1$, $\delta_{34}^{\rm S}=0$ :}& \multicolumn{1}{l}{$nsc(\bar{n}\bar{n})^{I=1}\bar{c}$,\quad $nsc\bar{s}\bar{s}\bar{c}$}\\
\bottomrule[0.1pt]
\multicolumn{2}{c}{\quad$\delta_{12}^{\rm A}=1$, $\delta_{34}^{\rm A}=0$, $\delta_{12}^{\rm S}=1$, $\delta_{34}^{\rm S}=1$ :}& \multicolumn{1}{l|}{$nsc(\bar{n}\bar{n})^{ I=0}\bar{c}$}&
\multicolumn{2}{c}{\quad$\delta_{12}^{\rm A}=1$, $\delta_{34}^{\rm A}=1$, $\delta_{12}^{\rm S}=1$, $\delta_{34}^{\rm S}=1$ :}& \multicolumn{1}{l}{$nsc\bar{n}\bar{s}\bar{c}$}\\
\bottomrule[0.5pt]
\bottomrule[1.5pt]
\end{tabular}
\end{lrbox}\scalebox{0.915}{\usebox{\tablebox}}
\end{table*}

\section{Numerical results and discussion}
\label{sec4}
Sandwiching the CMI Hamiltonian between the two wavefunctions with the same quantum number, we obtain the Hamiltonian matrices. Based on the corresponding eigenvalues and eigenvectors, we discuss the mass gaps, stabilities, and strong decay behaviors of all the hidden-charm hexaquark states.

\begin{figure*}[t]
\includegraphics[width=18.1cm]{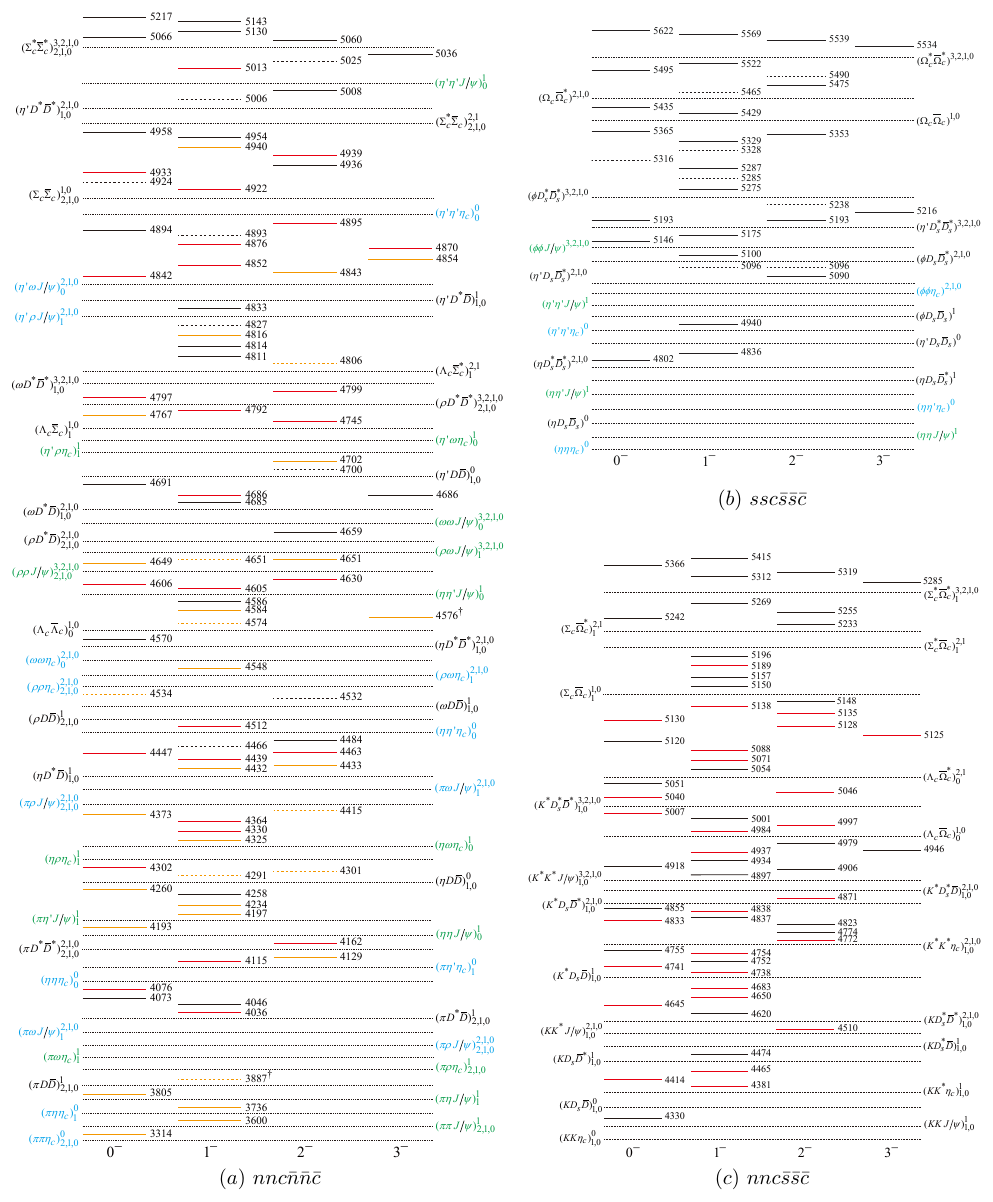}
\caption{Relative positions (units: MeV) for three kinds of hexaquark states.
	In the $nnc\bar{n}\bar{n}\bar{c}$ subsystem,
	the black lines show the $(nn)^{I=1}c(\bar{n}\bar{n})^{I=1}\bar{c}$ hexaquark states,
	the red lines show the $(nn)^{I=0}c(\bar{n}\bar{n})^{I=1}\bar{c}$ hexaquark states,
	and the orange lines show the $(nn)^{I=0}c(\bar{n}\bar{n})^{I=0}\bar{c}$ hexaquark states.
	The solid and dashed short lines are to differentiate the positive and negative $C$ parity and it's the same as $ssc\bar{s}\bar{s}\bar{c}$ subsystem.
	In the $nnc\bar{s}\bar{s}\bar{c}$ subsystem, the black (red) lines represent the $nnc\bar{s}\bar{s}\bar{c}$ hexaquark states with $I=1 (0)$.
	The dotted lines denote various baryon-antibaryon or meson-meson-meson thresholds.
	Some meson-meson-meson thresholds have specific $C$ parity, we label thresholds which have positive (negative) $C$ parity with blue (green).
	When the spin (isospin) of an initial hexaquark state is equal to the number in the superscript (subscript) of a baryon-antibaryon (meson-meson-meson) state, it can decay into these state through $S$-wave.
	Moreover, the stable states are marked with ``$\dag$".
}\label{1}
\end{figure*}

\begin{figure*}[t]
	\begin{tabular}{cc}
		\includegraphics[width=235pt]{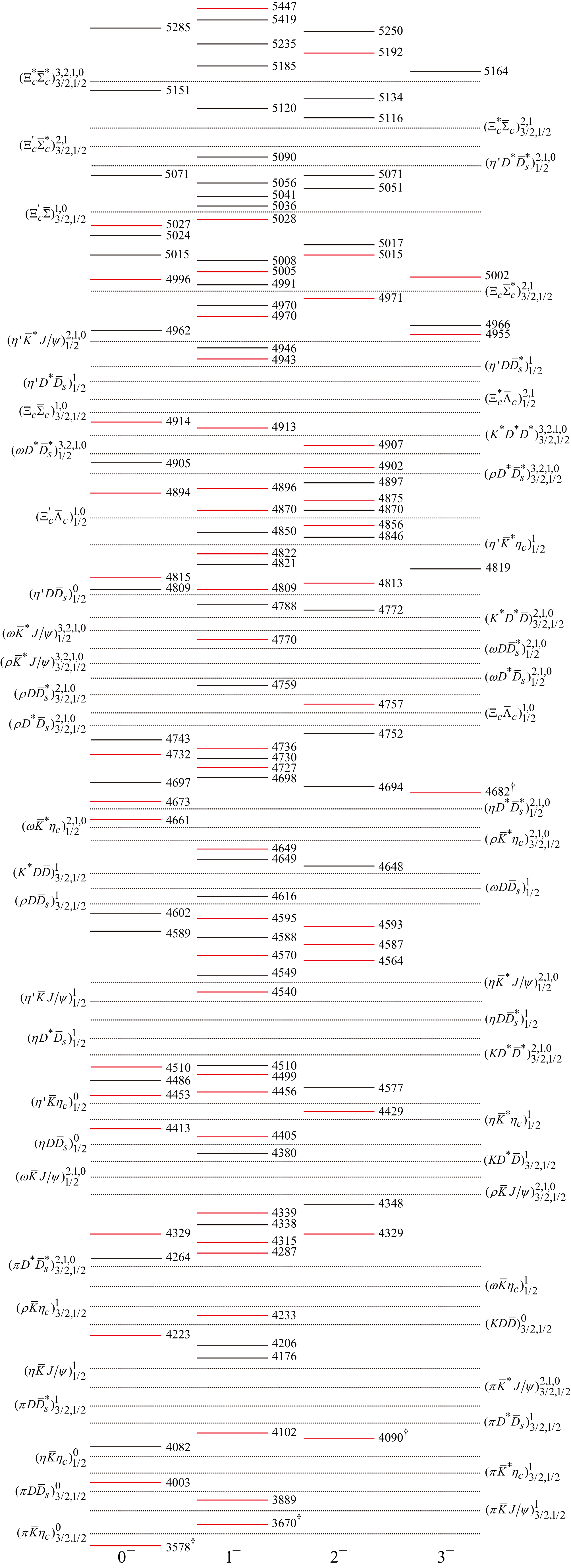}&
		\includegraphics[width=235pt]{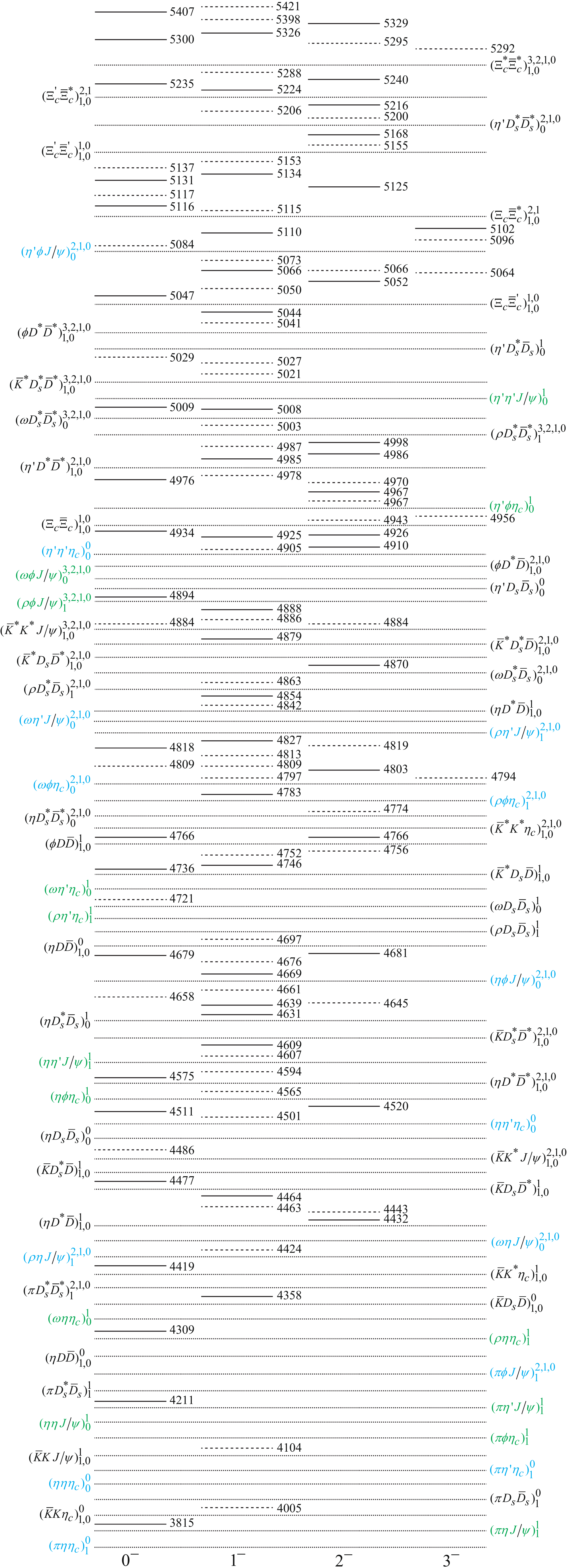}\\
		(a) \begin{tabular}{c}  $nsc\bar{n}\bar{n}\bar{c}$ states\end{tabular} &(b)  $nsc\bar{n}\bar{s}\bar{c}$ states\\
	\end{tabular}
	\caption{
		Relative positions (units: MeV) for the $nsc\bar{n}\bar{n}\bar{c}$ and $nsc\bar{n}\bar{s}\bar{c}$ hexaquark states labeled with solid lines.
		In the $nsc\bar{n}\bar{n}\bar{c}$ subsystem, the  black (red) lines represent the $nsc\bar{n}\bar{n}\bar{c}$ hexaquark states with $ I_{\bar{n}\bar{n}} =1 (0)$.
		In the $nsc\bar{n}\bar{s}\bar{c}$ subsystem, the solid (dashed) lines represent the $nsc\bar{n}\bar{s}\bar{c}$ hexaquark states with the positive (negative) $C$ parity.
		See the caption of Fig. \ref{1} for meaning of thresholds and ``$\dag$".
	}\label{2}
\end{figure*}

\begin{figure}[t]
\begin{tabular}{c}
\includegraphics[width=240pt]{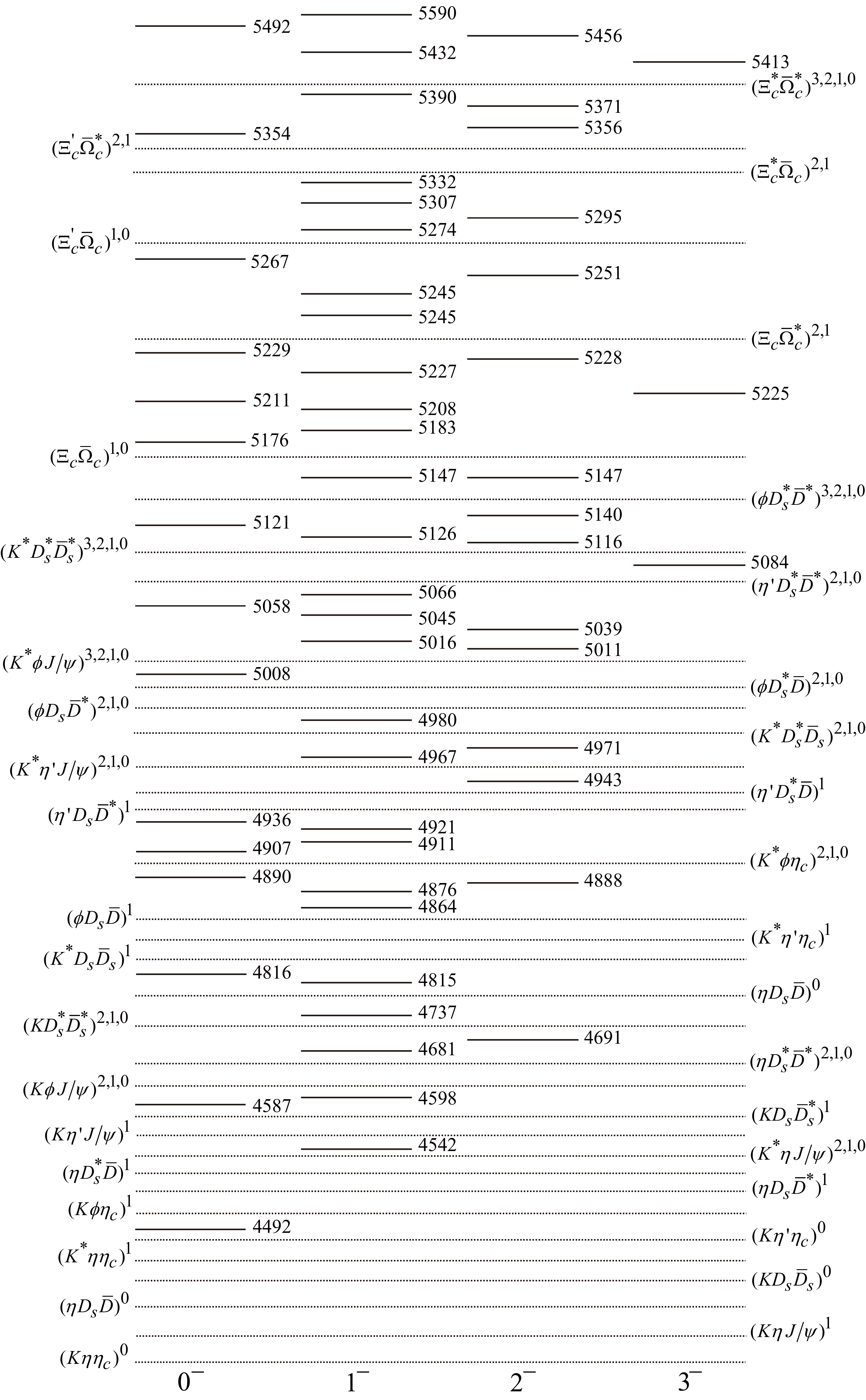}\\
(a)  $nsc\bar{s}\bar{s}\bar{c}$ states\\
\end{tabular}
\caption{
Relative positions (units: MeV) for the $nsc\bar{s}\bar{s}\bar{c}$ hexaquark states labeled with solid short lines.
The dotted lines denote various baryon-antibaryon and meson-meson-meson thresholds.
When the spin of an initial hexaquark state is equal to a number in the superscript of a baryon-antibaryon (meson-meson-meson) state, it can decay into these states through $S$-wave.
}\label{3}
\end{figure}

From the eigenvalues, we present the mass spectra in Fig. \ref{1} (for $nnc\bar{n}\bar{n}\bar{c}$, $ssc\bar{s}\bar{s}\bar{c}$, and
$nnc\bar{s}\bar{s}\bar{c}$), Fig. \ref{2} (for $nnc\bar{n}\bar{s}\bar{c}$ and $nsc\bar{n}\bar{s}\bar{c}$), and Fig. \ref{3} (for $nsc\bar{s}\bar{s}\bar{c}$). Moreover, we also plot the corresponding thresholds which they can decay to through quark rearrangements.
In convenience, we label the spin (isospin) of the rearrangement decay channel in the superscript (subscript).

In addition to the mass spectra, we discuss the two body strong decay based on the obtained eigenvectors. According to Table \ref{color} and Table \ref{type}, we can find that there are overlaps between hexaquark states and particular
baryon-antibaryon states. In the $qqc \otimes \bar{q}\bar{q}\bar{c}$ configuration, the color wavefunction of the hexaquark states falls into three categories:
$|(qqc)^{1_{c}}(\bar{q}\bar{q}\bar{c})^{1_{c}}\rangle$, $|(qqc)^{8_{c}}(\bar{q}\bar{q}\bar{c})^{8_{c}}\rangle$, and $|(qqc)^{10_{c}}(\bar{q}\bar{q}\bar{c})^{\bar{10}_{c}}\rangle$.
The $|(qqc)^{1_{c}}(\bar{q}\bar{q}\bar{c})^{1_{c}}\rangle$ can easily dissociate into an S-wave baryon and S-wave antibaryon (the ``OZI-superallowed" decay mode).
In contrast, the $|(qqc)^{8_{c}}(\bar{q}\bar{q}\bar{c})^{8_{c}}\rangle$ and $|(qqc)^{10_{c}}(\bar{q}\bar{q}\bar{c})^{\bar{10}_{c}}\rangle$ fall apart through the gluon exchange.
For simplicity, we only focus on the ``OZI-superallowed" decay mode.

The partial width of the two body $L$-wave ``OZI-superallowed decay" mode reads~\cite{Weng:2019ynva,Weng:2020jao,Weng:2021hje}
\begin{equation}
	\label{eqn:width}
	\Gamma_{i}=\gamma_{i}\alpha\frac{k^{2L+1}}{m^{2L}}{\cdot}|c_i|^2,
\end{equation}
where $\alpha$ is an effective coupling constant, $m$ is the initial state mass, $k$ is the spatial momentum of the final state in the center-of-mass frame, and $c_{i}$ is overlap between the hexaquark states and the final baryon-antibaryon states.
Generally, $\gamma_{i}$ depends on the spatial wavefunctions of the initial hexaquark and final baryon-antibaryon, which are different
for each decay process.
In the heavy quark limit, $\Sigma_{c}$ ($\Xi^{*}_{c}$) and $\Sigma^{*}_{c}$ ($\Xi'_{c}$) have the same spatial wavefunction.
Based on these, we assume the $\gamma_{i}$ relationships for different hidden-charm hexaquark states presented in Table~\ref{gamma}.
We find that the $(k/m)^2$ is of $\mathcal{O}(10^{-2})$ or even smaller, which means that the large partial wave decays are all suppressed.
Thus we only need to consider the $S$-wave two body decay modes.
Employing the eigenvectors in Table \ref{eig}, we calculate the values of $k\cdot|c_{i} |^{2}$ for each decay process and present them in Table \ref{value}.
The blank area in Tables \ref{eig} and \ref{value} means that the hexaquark state is forbidden to decay through this channel because of the quantum number conservation.
According to the $\gamma_{i}$ relationships in Table \ref{gamma} and the values of $k\cdot|c_{i} |^{2}$ in Table \ref{value},
we can roughly estimate the relative decay widths for different two-body decay processes of a hexaquark state.

\begin{table}[htp]
\caption{The $\gamma_{i}$ relationships for different hidden-charm hexaquark subsystems.
}\label{gamma}
\begin{center}
\renewcommand\arraystretch{1.4}
\begin{tabular}{c|l}
\toprule[1pt]
\toprule[0.5pt]
Subsystem&\multicolumn{1}{c}{$\gamma_{i}$}\\
\hline
\multirow{1}*{$nnc\bar{n}\bar{n}\bar{c}$}&$\gamma_{\Sigma_{c}^{*}\bar{\Sigma}_{c}^{*}}=\gamma_{\Sigma_{c}^{*}\bar{\Sigma}_{c}}=\gamma_{\Sigma_{c}\bar{\Sigma}^{*}_{c}}=\gamma_{\Sigma_{c}\bar{\Sigma}_{c}}$\quad$\gamma_{\Sigma_{c}^{*}\bar{\Lambda}_{c}}=\gamma_{\Sigma_{c}\bar{\Lambda}_{c}}$\\
\multirow{1}*{$nsc\bar{n}\bar{s}\bar{c}$}&$\gamma_{\Xi_{c}^{*} \bar{\Xi}^{*}_{c}}=\gamma_{\Xi _{c}' \bar{\Xi}_{c}^{*}}=\gamma_{\Xi_{c}' \bar{\Xi}_{c}'}
	=\gamma_{ \Xi _{c}\bar{\Xi} _{c}^{*}}\approx\gamma_{\Xi _{c}\bar{\Xi} _{c}'}=\gamma_{\Xi _{c} \bar{\Xi} _{c}}$\\
\multirow{1}*{$nsc\bar{n}\bar{n}\bar{c}$}&$\gamma_{\Xi _{c}^{*} \bar{\Sigma} _{c}^{*}}=\gamma_{\Xi _{c}^{*} \bar{\Sigma} _{c}}=\gamma_{\Xi _{c}' \bar{\Sigma} _{c}^{*}}=\gamma_{\Xi _{c}' \bar{\Sigma} _{c}}=\gamma_{\Xi _{c} \bar{\Sigma} _{c}^{*}}=\gamma_{\Xi _{c} \bar{\Sigma} _{c}}$\\
\multirow{1}*{$nnc\bar{s}\bar{s}\bar{c}$}&$\gamma_{\Sigma_c^{*}\bar{\Omega}_c^{*}}=\gamma_{\Sigma_c\bar{\Omega}_c^{*}}=\gamma_{\Sigma_c^{*}\bar{\Omega}_c}=\gamma_{\Sigma_c \bar{\Omega}_c}$\quad$\gamma_{\Lambda_c\bar{\Omega}_c^{*}}=\gamma_{\Lambda_c\bar{\Omega}_c}$\\
\multirow{1}*{$nsc\bar{s}\bar{s}\bar{c}$}&$\gamma_{\Xi_{c}^{*}\bar{\Omega}_{c}^{*}}=\gamma_{\Xi_{c}^{*}\bar{\Omega}_{c}}=\gamma_{\Xi_{c}^{'}\bar{\Omega}_{c}^{*}}=\gamma_{\Xi_{c}^{'}\bar{\Omega}_{c}}\approx\gamma_{\Xi_{c}\bar{\Omega}_{c}^{*}}=\gamma_{\Xi_{c}\bar{\Omega}_{c}}$\\
\multirow{1}*{$ssc\bar{s}\bar{s}\bar{c}$}&$\gamma_{\Omega_{c}^{*}\bar{\Omega}_{c}^{*}}= \gamma_{\Omega_{c}^{*}\bar{\Omega}_{c}}=\gamma_{\Omega_{c}^{*}\bar{\Omega}_{c}}= \gamma_{\Omega_{c}\bar{\Omega}_{c}}$\\
\toprule[0.5pt]
\toprule[1pt]
\end{tabular}
\end{center}
\end{table}

\begin{table*}[t]
	\centering \caption{
		The values of eigenvectors for the $nnc\bar{n}\bar{n}\bar{c}$, $nsc\bar{n}\bar{n}\bar{c}$, $nnc\bar{s}\bar{s}\bar{c}$, $nsc\bar{n}\bar{s}\bar{c}$, $nsc\bar{s}\bar{s}\bar{c}$, and $ssc\bar{s}\bar{s}\bar{c}$ hexaquark subsystems. The masses are all in units of MeV.
	}\label{eig}
	\begin{lrbox}{\tablebox}
		\renewcommand\arraystretch{1.19}
		\renewcommand\tabcolsep{1.55pt}
		\begin{tabular}{ccccccccccc|cccccccc|ccccc}
		\bottomrule[1.5pt]
		\bottomrule[0.5pt]
		\multicolumn{6}{l}{$ssc\bar{s}\bar{s}\bar{c}$\quad$(I=0)$}&\multicolumn{6}{|l}{$(nn)^{ I=1}c\bar{s}\bar{s}\bar{c}$\quad$(I=1)$}&\multicolumn{6}{|l}{$(nn)^{ I=1}c(\bar{n}\bar{n})^{ I=1}\bar{c}$\quad$(I=2,1,0)$}&&\multicolumn{5}{l}{$(nn)^{ I=0}c(\bar{n}\bar{n})^{ I=1}\bar{c}$\quad$(I=1)$}\\
		$J^{PC}$ & Mass & ${\Omega_{c}^{*}\bar{\Omega}_{c}^{*}}$& $\Omega_{c}^{*}\bar{\Omega}_{c}$&$\Omega_{c}\bar{\Omega}_{c}^{*}$&${\Omega_{c}\bar{\Omega}_{c}}$
		&\multicolumn{1}{|c}{$J^{P}$} & Mass& ${\Sigma_c^{*}\bar{\Omega}_c^{*}}$& ${\Sigma_c\bar{\Omega}_c^{*}}$& \multicolumn{1}{c}{${\Sigma_c^{*}\bar{\Omega}_c}$}& ${\Sigma_c \bar{\Omega}_c}$
		&\multicolumn{1}{|c}{$J^{PC}$}&Mass&$\Sigma^{*}_{c}\bar{\Sigma}^{*}_{c}$&$\Sigma^{*}_{c}\bar{\Sigma}_{c}$&$\Sigma_{c}\bar{\Sigma}^{*}_{c}$&$\Sigma_{c}\bar{\Sigma}_{c}$
		&&$J^{P}$&Mass&$\Lambda_{c}\bar{\Sigma}^{*}_{c}$&$\Lambda_{c}\bar{\Sigma}_{c}$&\\
		\bottomrule[0.5pt]

        $3^{--}$ & 5534 & -0.993 &  &  &  & \multicolumn{1}{|c}{$3^-$} & 5285 & -0.991 &  & \multicolumn{1}{c}{} &  & \multicolumn{1}{|c}{$3^{--}$} & 5036 & -0.989 &  &  &  &  & \multicolumn{1}{c}{$2^-$} & 4939 & 0.259 &  &  \\
        $2^{--}$ & 5490 &  & 0.665 & -0.665 &  & \multicolumn{1}{|c}{$2^-$} & 5319 & -0.884 & -0.306 & \multicolumn{1}{c}{-0.128} &  & \multicolumn{1}{|c}{$2^{--}$} & 5025 & 0.909 &  &  &  &  & \multicolumn{1}{c}{} & 4895 & -0.764 &  &  \\
        $2^{-+}$ & 5539 & -0.983 & -0.050 & -0.050 &  & \multicolumn{1}{|c}{} & 5255 & -0.288 & 0.358 & \multicolumn{1}{c}{0.720} &  & \multicolumn{1}{|c}{$2^{-+}$} & 5060 & -0.924 & -0.161 & -0.161 &  &  & \multicolumn{1}{c}{$1^-$} & 5013 & 0.221 & -0.015 &  \\
        & 5475 & 0.091 & -0.664 & -0.664 &  & \multicolumn{1}{|c}{} & 5233 & -0.259 & 0.825 & \multicolumn{1}{c}-0.441 &  & \multicolumn{1}{|c}{} & 5008 & 0.296 & -0.484 & -0.484 &  &  & \multicolumn{1}{c}{} & 4922 & 0.456 & 0.255 &  \\
        $1^{--}$ & 5569 & 0.918 & 0.135 & -0.135 & 0.133 & \multicolumn{1}{|c}{$1^-$} & 5415 & 0.688 & 0.3 & \multicolumn{1}{c}{0.275} & 0.188 & \multicolumn{1}{|c}{$1^{--}$} & 5143 & 0.672 & 0.327 & -0.327 & 0.338 &  & \multicolumn{1}{c}{} & 4876 & -0.634 & -0.300 &  \\
        & 5522 & -0.225 & 0.593 & -0.593 & 0.248 & \multicolumn{1}{|c}{} & 5312 & 0.417 & -0.116 & \multicolumn{1}{c}{-0.685} & -0.213 & \multicolumn{1}{|c}{} & 5130 & -0.538 & 0.416 & -0.416 & 0.161 &  & \multicolumn{1}{c}{} & 4852 & -0.354 & 0.740 &  \\
        & 5429 & 0.087 & 0.243 & -0.243 & -0.882 & \multicolumn{1}{|c}{} & 5269 & -0.307 & 0.805 & \multicolumn{1}{c}{-0.191} & -0.169 & \multicolumn{1}{|c}{} & 4954 & -0.217 & -0.415 & 0.415 & 0.825 &  & \multicolumn{1}{c}{} & 4792 & 0.128 & -0.264 &  \\
        $1^{-+}$ & 5465 &  & 0.699 & 0.699 &  & \multicolumn{1}{|c}{} & 5196 & -0.071 & 0.123 & \multicolumn{1}{c}{-0.439} & 0.774 & \multicolumn{1}{|c}{$1^{-+}$} & 5006 &  & 0.650 & 0.650 &  &  & \multicolumn{1}{c}{} & 4686 & 0.188 & -0.031 &  \\
        $0^{-+}$ & 5622 & -0.806 &  &  & -0.235 & \multicolumn{1}{|c}{} & 5157 & 0.058 & -0.173 & \multicolumn{1}{c}{-0.145} & -0.284 & \multicolumn{1}{|c}{$0^{-+}$} & 5217 & -0.704 &  &  & -0.369 &  & \multicolumn{1}{c}{} & 4605 & -0.191 & 0.097 &  \\
        & 5495 & 0.447 &  &  & -0.641 & \multicolumn{1}{|c}{} & 5150 & 0.308 & 0.244 & \multicolumn{1}{c}{0.067} & -0.227 & \multicolumn{1}{|c}{} & 5066 & 0.531 &  &  & -0.599 &  & \multicolumn{1}{c}{$0^-$} & 4933 &  & 0.118 &  \\
        & 5435 & -0.017 &  &  & 0.365 & \multicolumn{1}{|c}{$0^-$} & 5366 & -0.899 & \multicolumn{1}{l}{} & \multicolumn{1}{c}{} & -0.169 & \multicolumn{1}{|c}{} & 4958 & 0.055 &  &  & 0.213 &  & \multicolumn{1}{c}{} & 4842 &  & -0.599 &  \\
        &  &  &  &  &  & \multicolumn{1}{|c}{} & 5242 & 0.191 & \multicolumn{1}{l}{} & \multicolumn{1}{c}{} & -0.827 & \multicolumn{1}{|c}{} &  &  &  &  &  &  & \multicolumn{1}{c}{} & 4797 &  & 0.674 &  \\

		\bottomrule[1.0pt]
		\multicolumn{10}{l}{$nsc\bar{n}\bar{s}\bar{c}$\quad$(I=1,0)$}&\multicolumn{1}{c|}{}&
		\multicolumn{7}{l}{$nsc(\bar{n}\bar{n})^{ I=1}\bar{c}$\quad$(I=3/2,1/2)$}&\multicolumn{1}{c|}{}&
		\multicolumn{5}{l}{$(nn)^{I=0}c(\bar{n}\bar{n})^{I=0}\bar{c}$\quad$(I=0)$}\\
		$J^{PC}$ & Mass & $\Xi^{*}_{c}\bar{\Xi}^{*}_{c}$&$\Xi^{*}_{c}\bar{\Xi}'_{c}$&$\Xi'_{c}\bar{\Xi}^{*}_{c}$&$\Xi^{*}_{c}\bar{\Xi}_{c}$&$\Xi_{c}\bar{\Xi}^{*}_{c}$
		&$\Xi'_{c}\bar{\Xi}_{c}$&$\Xi_{c}\bar{\Xi}'_{c}$&$\Xi'_{c}\bar{\Xi}'_{c}$&$\Xi_{c}\bar{\Xi}_{c}$&
		$J^{P}$ & Mass & $\Xi^{*}_{c}\bar{\Sigma}^{*}_{c}$&$\Xi^{*}_{c}\bar{\Sigma}_{c}$&$\Xi'_{c}\bar{\Sigma}^{*}_{c}$&$\Xi'_{c}\bar{\Sigma}_{c}$&
		$\Xi_{c}\bar{\Sigma}^{*}_{c}$&$\Xi_{c}\bar{\Sigma}_{c}$&
		$J^{PC}$ & Mass & $\Lambda_{c}\bar{\Lambda}_{c}$ & & \\
		\bottomrule[0.5pt]
				
        $3^{--}$ & 5292 & -0.991 &  &  &  &  &  &  & & \multicolumn{1}{c|}{} & $3^-$ & 5164 & 0.989 &  &  &  &  &  & $1^{--}$ & 4940 & -0.319 &  &  \\
        $2^{-+}$ & 5329 & -0.883 & -0.148 & -0.148 & -0.132 & -0.132 &  & &  & \multicolumn{1}{c|}{} & $2^-$ & 5250 & 0.749 & 0.187 & 0.377 & 0.252 &  &  &  & 4816 & -0.019 &  &  \\
        & 5240 & 0.264 & -0.645 & -0.645 & -0.095 & -0.095 & &  &  & \multicolumn{1}{c|}{} &  & 5134 & -0.482 & 0.709 & 0.354 & 0.145 &  &  & &  4584 & -0.818 & &  \\
        & 5216 & -0.303 & -0.013 & -0.013 & 0.291 & 0.291 & &  &  & \multicolumn{1}{c|}{} &  & 5116 & -0.276 & -0.554 & 0.705 & -0.005 &  &  & $0^{-+}$ & 4767 & 0.562 & &  \\
        & 5168 & -0.064 & -0.168 & -0.168 & 0.535 & 0.535 & &  &  & \multicolumn{1}{c|}{} &  & 5071 & 0.203 & 0.140 & -0.013 & -0.707 &  &  &  & 4649 & -0.619 & &  \\

        \Xcline{20-24}{1pt}

        & 5125 & -0.030 & 0.081 & 0.081 & -0.006 & -0.006 &  &  &  &  &  & 5051 & -0.014 & 0.118 & 0.350 & -0.345 &  &  &

       \multicolumn{5}{l}{$nsc(\bar{n}\bar{n}\bar{c})^{I=0}$\quad$(I=1/2)$} \\
       $2^{--}$ & 5295 &  & 0.603 & -0.603 & 0.131 & -0.131 &  &  &  &  &  & 5017 & 0.080 & 0.180 & 0.156 & -0.293 &  &  & $J^{P}$ & Mass &$\Xi_{c}^{*}\bar{\Lambda}_{c}$&$\Xi_{c}'\bar{\Lambda}_{c}$&$\Xi_{c}\bar{\Lambda}_{c}$ \\

       \Xcline{20-24}{0.5pt}

       & 5200 &  & -0.026 & 0.026 & -0.443 & 0.443 & & &  &  & $1^-$ & 5419 & 0.509 & 0.341 & 0.303 & 0.231 & 0.259 & 0.225 & $2^-$ & 5192 & -0.061 &  &  \\
       & 5155 &  & 0.218 & -0.218 & -0.407 & 0.407 & &  &  &  &  & 5235 & -0.523 & 0.549 & -0.036 & 0.067 & 0.163 & 0.104 &  & 5015 & 0.746 &  &  \\
       $1^{-+}$ & 5326 &  & 0.347 & 0.347 & 0.194 & 0.194 & 0.219 & 0.219 &  &  &  & 5185 & -0.337 & -0.092 & 0.652 & 0.009 & -0.162 & -0.039 &  & 4971 & -0.359 &  &  \\
       & 5224 &  & -0.578 & -0.578 & 0.069 & 0.069 & 0.108 & 0.108 &  &  &  & 5120 & -0.18 & -0.042 & -0.302 & 0.366 & 0.206 & -0.011 & $1^-$ & 5447 & 0.037 & -0.023 & -0.010 \\
       & 5134 &  & 0.066 & 0.066 & -0.620 & -0.620 & 0.132 & 0.132 &  & &  & 5090 & -0.201 & -0.555 & 0.218 & 0.154 & 0.578 & 0.247 &  & 5028 & 0.409 & -0.060 & -0.009 \\
       & 5110 &  & 0.025 & 0.025 & -0.157 & -0.157 & -0.291 & -0.291 &  &  &  & 5056 & 0.016 & 0.172 & 0.119 & -0.591 & 0.389 & -0.102 &  & 5005 & 0.659 & 0.443 & 0.164 \\
       & 5066 &  & 0.101 & 0.101 & -0.003 & -0.003 & -0.440 & -0.440 &  &  &  & 5041 & 0.187 & -0.172 & -0.333 & -0.129 & 0.246 & -0.126 &  & 4970 & -0.143 & 0.345 & 0.066 \\
       $1^{--}$ & 5421 & 0.433 & 0.346 & -0.346 & 0.128 & -0.128 & 0.107 & -0.107 & 0.317 & 0.142 &  & 5036 & -0.175 & 0.022 & 0.033 & -0.189 & 0.299 & -0.25 &  & 4943 & -0.426 & 0.634 & 0.025 \\
       & 5398 & 0.558 & -0.053 & 0.053 & 0.066 & -0.066 & 0.082 & -0.082 & 0.066 & 0.025 &  & 5008 & 0.113 & 0.019 & 0.192 & 0.313 & 0.138 & -0.734 &  & 4913 & 0.028 & -0.033 & 0.078 \\
       & 5288 & -0.537 & 0.388 & -0.388 & 0.088 & -0.088 & 0.095 & -0.095 & 0.173 & 0.014 &  & 4991 & 0.101 & 0.002 & -0.008 & -0.302 & 0.014 & -0.052 &  & 4896 & 0.171 & 0.215 & 0.432 \\
       & 5206 & 0.240 & 0.361 & -0.361 & -0.131 & 0.131 & -0.092 & 0.092 & -0.734 & -0.079 &  & 4970 & 0.160 & -0.163 & 0.070 & 0.021 & -0.108 & -0.015 &  & 4870 & 0.097 & 0.009 & -0.588 \\
       & 5153 & 0.118 & 0.090 & -0.090 & -0.549 & 0.549 & -0.141 & 0.141 & 0.394 & -0.131 & & 4946 & -0.168 & 0.165 & -0.211 & 0.204 & 0.136 & -0.135 &  & 4822 & 0.017 & -0.014 & -0.279 \\
       & 5114 & -0.065 & -0.057 & 0.057 & -0.265 & 0.265 & 0.488 & -0.488 & -0.158 & 0.129 & $0^-$ & 5285 & 0.845 &  &  &  & 0.256 & 0.093 &  & 4809 & -0.226 & 0.017 & -0.272 \\
       & 5073 & 0.182 & -0.173 & 0.173 & 0.127 & -0.127 & 0.063 & -0.063 & -0.059 & -0.163 &  & 5151 & 0.333 &  &  &  & -0.723 & -0.049 &  & 4770 & 0.058 & 0.127 & 0.107 \\
       & 5050 & -0.167 & 0.008 & -0.008 & 0.109 & -0.109 & 0.023 & -0.023 & -0.092 & -0.259 &  & 5071 & -0.003 &  &  &  & 0.287 & 0.244 & $0^-$ & 5027 &  & 0.047 & -0.124 \\
       & 5041 & 0.008 & 0.004 & -0.004 & -0.001 & 0.001 & 0.008 & -0.008 & 0.007 & -0.054 &  & 5024 & 0.057 &  &  &  & 0.064 & -0.719 &  & 4996 &  & -0.544 & -0.224 \\
       & 5027 & 0.093 & 0.067 & -0.067 & 0.111 & -0.111 & 0.083 & -0.083 & 0.109 & -0.746 &  & 5015 & 0.018 &  &  &  & -0.272 & 0.167 &  & 4914 &  & -0.705 & 0.246 \\
       & 5021 & -0.009 & 0.028 & -0.028 & -0.006 & 0.006 & -0.082 & 0.082 & 0.065 & -0.038 &  & 4962 & 0.033 &  &  &  & 0.171 & -0.139 &  & 4894 &  & 0.087 & 0.639 \\
	
	   \Xcline{12-19}{1pt}
	
      & 5003 & 0.001 & -0.019 & 0.019 & -0.124 & 0.124 & 0.296 & -0.296 & -0.016 & -0.291 & \multicolumn{8}{l|}{$nsc\bar{s}\bar{s}\bar{c}$\quad$(I=1/2)$} & & 4815 &  & 0.145 & 0.296 \\

       \Xcline{20-24}{1pt}

       & 4987 & -0.059 & -0.030 & 0.030 & 0.046 & -0.046 & 0.127 & -0.127 & -0.042 & 0.148
       &$J^{P}$&Mass&${\Xi_{c}^{*}\bar{\Omega}_{c}^{*}}$ &${\Xi_{c}'\bar{\Omega}_{c}^{*}}$&${\Xi_{c}^{*}\bar{\Omega}_{c}}$&${\Xi_{c}' \bar{\Omega}_{c}}$&${\Xi_{c}\bar{\Omega}_{c}^{*}}$&${\Xi_{c}\bar{\Omega}_{c}}$

       &\multicolumn{5}{l}{$(nnc)^{I=0}\bar{s}\bar{s}\bar{c}$\quad$(I=0)$} \\
	
       \Xcline{12-19}{0.5pt}

       & 4978 & 0.047 & 0.028 & -0.028 & -0.002 & 0.002 & -0.227 & 0.227 & 0.029 & -0.049 & $3^-$ & 5413 & 0.991  & & & & & & $J^P$ & Mass &

        $\Lambda_{c}\bar{\Omega}_{c}^{*}$ & $\Lambda_{c}\bar{\Omega}_{c}$ &    \\

	   \Xcline{20-24}{0.5pt}
	
      $0^{-+}$ & 5407 & -0.854 &  &  &  &  & -0.094 & -0.094 & -0.233 & -0.112 & $1^-$ & 5590 & -0.556 & -0.295 & -0.326 & -0.239 & -0.213 & -0.199 & $2^-$ & 5135 & 0.412 &  &  \\
      & 5300 & -0.064 &  &  &  &  & -0.052 & -0.052 & 0.342 & 0.027 &  & 5432 & 0.642 & -0.16 & -0.538 & -0.162 & -0.109 & -0.103 &  & 5128 & -0.661 &  &  \\
      & 5235 & 0.323 &  &  &  &  & -0.165 & -0.165 & -0.732 & -0.137 &  & 5390 & 0.192 & -0.776 & 0.326 & 0.149 & 0.015 & 0.033 & $1^-$ & 5189 & -0.337 & -0.068 &  \\
      & 5131 & 0.094 &  &  &  &  & -0.541 & -0.541 & 0.309 & -0.177 &  & 5332 & 0.178 & 0.102 & 0.502 & -0.560 & -0.408 & -0.181 &  & 5138 & -0.478 & -0.225 &  \\
      & 5116 & 0.018 &  &  &  &  & 0.030 & 0.030 & -0.016 & -0.264 &  & 5307 & -0.003 & -0.202 & 0.184 & -0.474 & 0.454 & -0.041 &  & 5088 & 0.677 & -0.0003 &  \\
      & 5047 & -0.071 &  &  &  &  & -0.202 & -0.202 & -0.175 & 0.733 &  & 5274 & 0.015 & 0.068 & -0.176 & -0.386 & 0.472 & 0.205 &  & 5071 & 0.139 & -0.819 &  \\
      & 5009 & 0.019 &  &  &  &  & -0.216 & -0.216 & 0.092 & -0.042 &  & 5245 & -0.001 & 0.106 & 0.068 & -0.138 & -0.297 & 0.634 &  & 4984 & 0.006 & -0.195 &  \\
      & 4976 & 0.041 &  &  &  &  & -0.052 & -0.052 & -0.035 & 0.167 &  & 5244 & 0.199 & -0.119 & -0.113 & -0.125 & -0.112 & 0.129 & $0^-$ & 5130 &  & 0.205 &  \\
      $0^{--}$ & 5137 &  &  &  &  &  & 0.280 & -0.280 & &  &  & 5227 & -0.166 & -0.298 & -0.12 & -0.064 & -0.166 & 0.443 &  & 5040 &  & 0.818 &  \\
      & 5117 &  &  &  &  &  & 0.382 & -0.382 &  &  &  & 5208 & -0.093 & 0.039 & -0.012 & -0.091 & 0.260 & 0.182 &  & 5007 &  & 0.359 &  \\

      \Xcline{20-24}{1pt}

      & 5084 &  &  &  &  &  & -0.441 & 0.441 &  &  &  & 5183 & -0.141 & -0.151 & 0.186 & 0.148 & 0.116 & -0.021 & $J^{P}$ & Mass & $\Xi_{c}' \bar{\Omega}_{c}^{*}$ & $\Xi_{c}' \bar{\Omega}_{c}$ &$\Xi_{c} \bar{\Omega}_{c}$ \\

	  \Xcline{1-11}{1pt}
	  \Xcline{20-24}{0.5pt}

	  $J^{P}$ & Mass & $\Xi_{c}^{*} \bar{\Omega}_{c}^{*}$ & $\Xi_{c}' \bar{\Omega}_{c}^{*}$ &$\Xi_{c}^{*} \bar{\Omega}_{c}$ &$\Xi_{c}^{*} \bar{\Omega}_{c}$& Mass&$\Xi_{c}^{*} \bar{\Omega}_{c}^{*}$ & $\Xi_{c}' \bar{\Omega}_{c}^{*}$ &$\Xi_{c}^{*} \bar{\Omega}_{c}$ &$\Xi_{c}^{*} \bar{\Omega}_{c}$ & $0^-$ & 5492 & 0.899 &  &  & 0.171 &  &  \multicolumn{1}{c}{0.072}  &  $0^-$ & 5211 & -0.054 &  -0.069 &  -0.071
	   \\
	
	  \Xcline{1-11}{0.5pt}
	
	   $2^-$ & 5456 & 0.843 & 0.314 & 0.155 & 0.187 &  5295 & -0.16 & -0.168 & -0.126 & \multicolumn{1}{c}{0.855} & & 5354 & -0.216 &  &  & 0.829 &  & \multicolumn{1}{c}{0.071} &  & 5176 & -0.035 &   0.239 &  -0.333  \\
       & 5371 & -0.437 & 0.629 & 0.551 & 0.147 & 5251 & 0.057 & -0.134 & -0.214 & \multicolumn{1}{c}{0.123}  & & 5267 & 0.023 &  &  & -0.199 &  & \multicolumn{1}{c}{-0.411} &  &  &  &  &  \\

       & 5356 & 0.113 & -0.611 & 0.735 & 0.0003 &  & &  & & \multicolumn{1}{c}{} & & 5229 & 0.085 &  &  & 0.162 &  & \multicolumn{1}{c}{-0.710} &  &  &  &  &  \\
		\bottomrule[0.5pt]
		\bottomrule[1.5pt]	
		\end{tabular}
	\end{lrbox}\scalebox{0.81}{\usebox{\tablebox}}
\end{table*}

\begin{table*}[htbp]
	\centering \caption{
	The values of $k\cdot |c_{i}|^{2}$ for the $nnc\bar{n}\bar{n}\bar{c}$, $nsc\bar{n}\bar{n}\bar{c}$, $nnc\bar{s}\bar{s}\bar{c}$, $nsc\bar{n}\bar{s}\bar{c}$, $nsc\bar{s}\bar{s}\bar{c}$, and $ssc\bar{s}\bar{s}\bar{c}$ hexaquark subsystems. The masses are all in units of MeV. The $\times$ means that the decay channel is kinetically forbidden.}\label{value}
	\begin{lrbox}{\tablebox}
	\renewcommand\arraystretch{1.19}
	\renewcommand\tabcolsep{1.55pt}
	\begin{tabular}{ccccccccccc|cccccccc|ccccc}
		\bottomrule[1.5pt]
		\bottomrule[0.5pt]
		\multicolumn{6}{l}{$ssc\bar{s}\bar{s}\bar{c}$\quad$(I=0)$}&\multicolumn{6}{|l}{$(nn)^{ I=1}c\bar{s}\bar{s}\bar{c}$\quad$(I=1)$}&\multicolumn{6}{|l}{$(nn)^{I=1}c(\bar{n}\bar{n})^{I=1}\bar{c}$\quad$(I=2,1,0)$}&&\multicolumn{5}{l}{$(nn)^{I=0}c(\bar{n}\bar{n})^{I=1}\bar{c}$\quad$(I=1)$}\\
		$J^{PC}$ & Mass & ${\Omega_{c}^{*}\bar{\Omega}_{c}^{*}}$& $\Omega_{c}^{*}\bar{\Omega}_{c}$&$\Omega_{c}\bar{\Omega}_{c}^{*}$&${\Omega_{c}\bar{\Omega}_{c}}$
		&\multicolumn{1}{|c}{$J^{P}$} & Mass& ${\Sigma_c^{*}\bar{\Omega}_c^{*}}$& ${\Sigma_c\bar{\Omega}_c^{*}}$& \multicolumn{1}{c}{${\Sigma_c^{*}\bar{\Omega}_c}$}& ${\Sigma_c\bar{\Omega}_c}$
		&\multicolumn{1}{|c}{$J^{PC}$}&Mass&$\Sigma^{*}_{c}\bar{\Sigma}^{*}_{c}$&$\Sigma^{*}_{c}\bar{\Sigma}_{c}$&$\Sigma_{c}\bar{\Sigma}^{*}_{c}$&$\Sigma_{c}\bar{\Sigma}_{c}$
		&&$J^{P}$&Mass&$\Lambda_{c}\bar{\Sigma}^{*}_{c}$&$\Lambda_{c}\bar{\Sigma}_{c}$&\\
		\bottomrule[0.5pt]
		
		$3^{--}$ & 5534 & 69 &  &  &  & \multicolumn{1}{|c}{$3^-$} & 5285 & 47 &  & \multicolumn{1}{c}{} &  & \multicolumn{1}{|c}{$3^{--}$} & 5036 & 552 &  &  &  &  & \multicolumn{1}{c}{$2^-$} & 4939 & 38 &  &  \\
		$2^{--}$ & 5490 &  & 124 & 124 &  & \multicolumn{1}{|c}{$2^-$} & 5319 & 238 & 48 & \multicolumn{1}{c}{9} &  & \multicolumn{1}{|c}{$2^{--}$} & 5025 & 301 &  &  &  &  & \multicolumn{1}{c}{} & 4895 & 272 &  &  \\
		$2^{-+}$ & 5539 & 136 & 1 & 1 &  & \multicolumn{1}{|c}{} & 5255 & $\times$ & 38.5 & \multicolumn{1}{c}{170} &  & \multicolumn{1}{|c}{$2^{-+}$} & 5060 & 206 & 12 & 12 &  &  & \multicolumn{1}{c}{$1^-$} & 5013 & 35 & 0.2 &  \\
		& 5475 & $\times$ & 88 & 88 &  & \multicolumn{1}{|c}{} & 5233 & $\times$ & 126 & \multicolumn{1}{c}{44} &  & \multicolumn{1}{|c}{} & 5008 & $\times$ & 70 & 70 &  &  & \multicolumn{1}{c}{} & 4922 & 110 & 59 &  \\
		$1^{--}$ & 5569 & 271 & 10 & 10 & 12 & \multicolumn{1}{|c}{$1^-$} & 5415 & 281 & 65 & \multicolumn{1}{c}{56} & 30 & \multicolumn{1}{|c}{$1^{--}$} & 5143 & 235 & 71 & 71 & 88 &  & \multicolumn{1}{c}{} & 4876 & 167 & 76 &  \\
		& 5522 & $\times$ & 143 & 143 & 37 & \multicolumn{1}{|c}{} & 5312 & 47 & 7 & \multicolumn{1}{c}{239} & 30 & \multicolumn{1}{|c}{} & 5130 & 141 & 109 & 109 & 19 &  & \multicolumn{1}{c}{} & 4852 & 42 & 444 &  \\
		& 5429 & $\times$ & $\times$ & $\times$ & 252 & \multicolumn{1}{|c}{} & 5269 & $\times$ & 233 & \multicolumn{1}{c}{14} & 16 & \multicolumn{1}{|c}{} & 4954 & $\times$ & $\times$ & $\times$ & 228 &  & \multicolumn{1}{c}{} & 4792 & $\times$ & 50 &  \\
		$1^{-+}$ & 5465 &  & 53 & 53 &  & \multicolumn{1}{|c}{} & 5196 & $\times$ & $\times$ & \multicolumn{1}{c}{} & 209 & \multicolumn{1}{|c}{$1^{-+}$} & 5006 &  & 123 & 123 &  &  & \multicolumn{1}{c}{} & 4686 & $\times$ & 0.5 &  \\
		$0^{-+}$ & 5622 & 325 &  &  & 44 & \multicolumn{1}{|c}{} & 5157 & $\times$ & $\times$ & \multicolumn{1}{c}{} & 11 & \multicolumn{1}{|c}{$0^{-+}$} & 5217 & 337 &  &  & 121 &  & \multicolumn{1}{c}{} & 4605 & $\times$ & 3 &  \\
		& 5495 & $\times$ &  &  & 219 & \multicolumn{1}{|c}{} & 5150 & $\times$ & $\times$ & \multicolumn{1}{c}{} & 2 & \multicolumn{1}{|c}{} & 5066 & 77 &  &  & 226 &  & \multicolumn{1}{c}{$0^-$} & 4933 &  & 10 &  \\
		& 5435 & $\times$ &  &  & 46 & \multicolumn{1}{|c}{$0^-$} & 5366 & 377 & \multicolumn{1}{l}{} & \multicolumn{1}{c}{} & 21 & \multicolumn{1}{|c}{} & 4958 & $\times$ &  &  & 16 &  & \multicolumn{1}{c}{} & 4842 &  & 178 &  \\
		&  &  &  &  &  & \multicolumn{1}{|c}{} & 5242 & $\times$ & \multicolumn{1}{l}{} & \multicolumn{1}{c}{} & 336 & \multicolumn{1}{|c}{} &  &  &  &  &  &  & \multicolumn{1}{c}{} & 4797 &  & 166 &   \\

		\bottomrule[1.0pt]
		\multicolumn{10}{l}{$nsc\bar{n}\bar{s}\bar{c}$\quad$(I=1,0)$}&\multicolumn{1}{c|}{}&
		\multicolumn{7}{l}{$nsc(\bar{n}\bar{n})^{I=1}\bar{c}$\quad$(I=3/2,1/2)$}&\multicolumn{1}{c|}{}&
		\multicolumn{5}{l}{$(nn)^{I=0}c(\bar{n}\bar{n})^{I=0}\bar{c}$\quad$(I=0)$}\\
		$J^{PC}$ & Mass & $\Xi^{*}_{c}\bar{\Xi}^{*}_{c}$&$\Xi^{*}_{c}\bar{\Xi}'_{c}$&$\Xi'_{c}\bar{\Xi}^{*}_{c}$&$\Xi^{*}_{c}\bar{\Xi}_{c}$&$\Xi_{c}\bar{\Xi}^{*}_{c}$
		&$\Xi'_{c}\bar{\Xi}_{c}$&$\Xi_{c}\bar{\Xi}'_{c}$&$\Xi'_{c}\bar{\Xi}'_{c}$&$\Xi_{c}\bar{\Xi}_{c}$&
		$J^{P}$ & Mass & $\Xi^{*}_{c}\bar{\Sigma}^{*}_{c}$&$\Xi^{*}_{c}\bar{\Sigma}_{c}$&$\Xi'_{c}\bar{\Sigma}^{*}_{c}$&$\Xi'_{c}\bar{\Sigma}_{c}$&
		$\Xi_{c}\bar{\Sigma}^{*}_{c}$&$\Xi_{c}\bar{\Sigma}_{c}$&
		$J^{PC}$ & Mass & $\Lambda_{c}\bar{\Lambda}_{c}$ & & \\
		\bottomrule[0.5pt]
		
		$3^{--}$ & 5292 & 51.2 &  &  &  &  &  &  &  & \multicolumn{1}{c|}{} & $3^-$ & 5164 & 28.4 &  &  &  &  &  & $1^{--}$ & 4940 & 95 &  &  \\
		$2^{-+}$ & 5329 & 247 & 12 & 12 & 13 & 13 &  &  &  & \multicolumn{1}{c|}{} & $2^-$ & 5250 & 265 & 22 & 90 & 52 &  &  &  & 4816 & 0.3 & &  \\
		& 5240 & $\times$ & 88 & 88 & 5 & 5 &  & &  & \multicolumn{1}{c|}{} &  & 5134 & $\times$ & 149 & 39 & 13 &  &  & & 4584 & 108 & &  \\
		& 5216 & $\times$ & $\times$ & $\times$ & 44 & 44 &  &  &  & \multicolumn{1}{c|}{} &  & 5116 & $\times$ & 64 & 114 & 0.01 &  &  & $0^{-+}$ & 4767 & 213  & & \\
		& 5168 & $\times$ & $\times$ & $\times$ & 107 & 107 & &  &  & \multicolumn{1}{c|}{} &  & 5071 & $\times$ & $\times$ & $\times$ & 231 & &  &  & 4649 & 160 & &   \\
		
		\Xcline{20-24}{1pt}
		
		 & 5125 & $\times$ & $\times$ & $\times$ & 0 & 0 &  &  &  & \multicolumn{1}{c|}{} &  & 5051 & $\times$ & $\times$ & $\times$ & 48 &  &  &
		\multicolumn{5}{l}{$nsc(\bar{n}\bar{n})^{I=0}\bar{c}$\quad$(I=1/2)$} \\
		
		$2^{--}$ & 5295 &  & 317 & 318 & 12 & 12 & &  &  & \multicolumn{1}{c|}{} &  & 5017 & $\times$ & $\times$ & $\times$ & 24 &  &  & $J^{P}$ & Mass &$\Xi_{c}^{*}\bar{\Lambda}_{c}$&$\Xi_{c}'\bar{\Lambda}_{c}$&$\Xi_{c}\bar{\Lambda}_{c}$ \\
		
		\Xcline{20-24}{0.5pt}
		
		& 5200 &  & $\times$ & $\times$ & 93 & 93 &  &  &  &  & $1^-$ & 5419 & 213 & 106 & 85 & 57 & 68 & 57 & $2^-$ & 5192 & 3 &  &  \\
		& 5155 &  & $\times$ & $\times$ & 54 & 54 &  & &  &  &  & 5235 & 118 & 179 & 1 & 4 & 19 & 10 &  & 5015 & 252 &  &  \\
		$1^{-+}$ & 5326 &  & 63 & 63 & 28 & 28 & 41 & 41 &  &  &  & 5185 & 27 & 4 & 204 & 0.1 & 16 & 1 &  & 4971 & 40 &  &  \\
		& 5224 &  & 17 & 17 & 3 & 3 & 8 & 8 &  &  &  & 5120 & $\times$ & 0.4 & 23 & 78 & 20 & 0.1 & $1^-$ & 5447 & 2 & 1 & 0.1 \\
		& 5134 &  & $\times$ & $\times$ & 88 & 88 & 9 & 9 &  &  &  & 5090 & $\times$ & $\times$ & $\times$ & 12 & 128 & 40 &  & 5028 & 82 & 2 & 0.1 \\
		& 5110 &  & $\times$ & $\times$ & $\times$ & $\times$ & 35 & 35 &  &  &  & 5056 & $\times$ & $\times$ & $\times$ & 146 & 38 & 6 &  & 5005 & 185 & 116 & 21 \\
		& 5066 &  & $\times$ & $\times$ & $\times$ & $\times$ & 44 & 44 &  &  &  & 5041 & $\times$ & $\times$ & $\times$ & 6 & 9 & 9 &  & 4970 & 6 & 61 & 3 \\
		$1^{--}$ & 5421 & 111 & 88 & 88 & 15 & 15 & 12 & 12 & 84 & 23 &  & 5036 & $\times$ & $\times$ & $\times$ & 13 & 9 & 33 &  & 4943 & 30 & 177 & 0.4 \\
		& 5398 & 167 & 2 & 2 & 4 & 4 & 7 & 7 & 4 & 1 &  & 5008 & $\times$ & $\times$ & $\times$ & 23 & $\times$ & 249 &  & 4913 & $\times$ & 0.4 & 4 \\
		& 5288 & $\times$ & 63 & 63 & 6 & 6 & 7 & 7 & 18 & 0.2 &  & 4991 & $\times$ & $\times$ & $\times$ & 9 & $\times$ & 1 &  & 4896 & $\times$ & 13 & 109 \\
		& 5206 & $\times$ & $\times$ & $\times$ & 9 & 9 & 6 & 6 & 196 & 5 &  & 4970 & $\times$ & $\times$ & $\times$ & $\times$ & $\times$ & 0.1 &  & 4870 & $\times$ & 0.01 & 182 \\
		& 5153 & $\times$ & $\times$ & $\times$ & 96 & 96 & 11 & 11 & $\times$ & 13 &  & 4946 & $\times$ & $\times$ & $\times$ & $\times$ & $\times$ & 4 &  & 4822 & $\times$ & $\times$ & 31 \\
		& 5114 & $\times$ & $\times$ & $\times$ & 4 & 4 & 100 & 100 & $\times$ & 11 & $0^-$ & 5285 & 402 &  &  &  & 53 & 8 &  & 4809 & $\times$ & $\times$ & 27 \\
		& 5073 & $\times$ & $\times$ & $\times$ & $\times$ & $\times$ & 1 & 1 & $\times$ & 16 &  & 5151 & $\times$ &  &  &  & 288 & 2 &  & 4770 & $\times$ & $\times$ & 2 \\
		& 5050 & $\times$ & $\times$ & $\times$ & $\times$ & $\times$ & 0 & 0 & $\times$ & 36 &  & 5071 & $\times$ &  &  &  & 26 & 36 & $0^-$ & 5027 &  & 1 & 13 \\
		& 5041 & $\times$ & $\times$ & $\times$ & $\times$ & $\times$ & $\times$ & $\times$ & $\times$ & 2 &  & 5024 & $\times$ &  &  & & $\times$ & 261 &  & 4996 &  & 169 & 38 \\
		& 5027 & $\times$ & $\times$ & $\times$ & $\times$ & $\times$ & $\times$ & $\times$ & $\times$ & 265 &  & 5015 & $\times$ &  &  &  & $\times$ & 13 &  & 4914 &  & 173 & 38 \\
		& 5021 & $\times$ & $\times$ & $\times$ & $\times$ & $\times$ & $\times$ & $\times$ & $\times$ & 1 &  & 4962 & $\times$ &  &  &  & $\times$ & 6 &  & 4894 &  & 2 & 237 \\
		
		\Xcline{12-19}{1pt}
		
		& 5003 & $\times$ & $\times$ & $\times$ & $\times$ & $\times$ & $\times$ & $\times$ & $\times$ & 35  & \multicolumn{8}{l|}{$nsc\bar{s}\bar{s}\bar{c}$\quad$(I=1/2)$} & & 4815 &  & $\times$ &  33 \\
		
		\Xcline{20-24}{1pt}
		
		& 4987 & $\times$ & $\times$ & $\times$ & $\times$ & $\times$ & $\times$ & $\times$ & $\times$ & 8 &$J^{P}$&Mass&${\Xi_{c}^{*}\bar{\Omega}_{c}^{*}}$ &${\Xi_{c}'\bar{\Omega}_{c}^{*}}$&${\Xi_{c}^{*}\bar{\Omega}_{c}}$&${\Xi_{c}' \bar{\Omega}_{c}}$&${\Xi_{c}\bar{\Omega}_{c}^{*}}$&${\Xi_{c}\bar{\Omega}_{c}}$
		&\multicolumn{5}{l}{$(nn)^{I=0}c\bar{s}\bar{s}\bar{c}$\quad$(I=0)$} \\
		
		\Xcline{12-19}{0.5pt}
		
		& 4978 & $\times$ & $\times$ & $\times$ & $\times$ & $\times$ & $\times$ & $\times$ & $\times$ & 1 & $3^-$ & 5413 & 68  &  &  &  &  &  & $J^P$ & Mass &
		$\Lambda_{c}\bar{\Omega}_{c}^{*}$ & $\Lambda_{c}\bar{\Omega}_{c}$ &    \\
		
		\Xcline{20-24}{0.5pt}
		
		$0^{-+}$ & 5407 & 407 &  &  &  &  & 9 & 9 & 44 & 14 & $1^-$ & 5590 & 216 & 71 & 88 & 53 & 45 & 43 & $2^-$ & 5135 & 78 &  &  \\
		& 5300 & 1 &  &  &  &  & 2 & 2 & 72 & 1 &  & 5432 & 98 & 13 & 144 & 17 & 9 & 9 &  & 5128 & 191 &  &  \\
		& 5235 & $\times$ &  &  &  &  & 19 & 19 & 245 & 16 &  & 5390 & $\times$ & 212 & 39 & 13 & 0.2 & 1 & $1^-$ & 5189 & 67 & 3 &  \\
		& 5131 & $\times$ &  &  &  &  & 136 & 136 & $\times$ & 22 &  & 5332 & $\times$ & $\times$ & $\times$ & 124 & 84 & 22 &  & 5138 & 106 & 32 &  \\
		& 5116 & $\times$ &  &  &  &  & 0.5 & 0.5 & $\times$ & 47 &  & 5307 & $\times$ & $\times$ & $\times$ & 68 & 91 & 1 &  & 5088 & 136 & 0.00005 &  \\
		& 5047 & $\times$ &  &  &  &  & 0.5 & 0.5 & $\times$ & 283 &  & 5274 & $\times$ & $\times$ & $\times$ & 9 & 72 & 23 &  & 5071 & 4 & 318 &  \\
		& 5009 & $\times$ &  &  &  &  & 3 & 3 & $\times$ & 1 &  & 5245 & $\times$ & $\times$ & $\times$ & $\times$ & 15 & 185 &  & 4984 & $\times$ & 3 &  \\
		& 4976 & $\times$ &  &  &  &  & $\times$ & $\times$ & $\times$ & 9 &  & 5244 & $\times$ & $\times$ & $\times$ & $\times$ & 2 & 8 & $0^-$ & 5130 & $\times$ & 26 &  \\
		$0^{--}$ & 5137 &  &  &  &  &  & 38 & 38 &  &  &  & 5227 & $\times$ & $\times$ & $\times$ & $\times$ & $\times$ & 80 &  & 5040 & $\times$ & 256 &  \\
		& 5117 &  &  &  &  &  & 62 & 62 &  &  &  & 5208 & $\times$ & $\times$ & $\times$ & $\times$ & $\times$ & 11 &  & 5007 & $\times$ & 33 &  \\
		
		\Xcline{20-24}{1pt}
		
		& 5084 &  &  &  &  &  & 62 & 62 &  &  &  & 5183 & $\times$ & $\times$ & $\times$ & $\times$ & $\times$ & 0.1 & $J^{P}$ & Mass & $\Xi_{c}' \bar{\Omega}_{c}^{*}$ & $\Xi_{c}' \bar{\Omega}_{c}$ &$\Xi_{c} \bar{\Omega}_{c}$ \\
		
		\Xcline{1-11}{1pt}
		\Xcline{20-24}{0.5pt}

		$J^{P}$ & Mass & $\Xi_{c}^{*} \bar{\Omega}_{c}^{*}$ & $\Xi_{c}' \bar{\Omega}_{c}^{*}$ &$\Xi_{c}^{*} \bar{\Omega}_{c}$ &$\Xi_{c}^{*} \bar{\Omega}_{c}$& Mass&$\Xi_{c}^{*} \bar{\Omega}_{c}^{*}$ & $\Xi_{c}' \bar{\Omega}_{c}^{*}$ &$\Xi_{c}^{*} \bar{\Omega}_{c}$ &$\Xi_{c}^{*} \bar{\Omega}_{c}$ & $0^-$ & 5492 & 1&  &  & 0.2 &  &  \multicolumn{1}{c}{0.1}  &  $0^-$ & 5211 & $\times$ & $\times$ & 2
		\\
		
		\Xcline{1-11}{0.5pt}
		
       $2^-$ & 5456 & 249 & 55 & 13 & 27 & 5295 & $\times$ & $\times$ & $\times$ & \multicolumn{1}{c}{294} &  & 5354 & $\times$ &  &  & 319 &  & \multicolumn{1}{c}{4} &  & 5176 & $\times$ & $\times$ & 21 \\
	   & 5371 & $\times$ & 108 & 86 & 13 & 5251 & $\times$ & $\times$ & $\times$ & \multicolumn{1}{c}{3} &  & 5267 & $\times$ &  &  & $\times$ &  & \multicolumn{1}{c}{88} &  & & & & \\
       & 5356 & $\times$ & 70 & 111 & 0 & & &  &  & \multicolumn{1}{c}{} &  & 5229 & $\times$ &  &  & $\times$ &  & \multicolumn{1}{c}{209} &  &  &  &  &  \\
		\bottomrule[0.5pt]
		\bottomrule[1.5pt]	
	\end{tabular}
\end{lrbox}\scalebox{0.83}{\usebox{\tablebox}}
\end{table*}

\subsection{The $nnc\bar{n}\bar{n}\bar{c}$ subsystem}
\label{sec:nnc.n.n.c}

Firstly, we discuss the $nnc\bar{n}\bar{n}\bar{c}$ subsystem based on Fig. \ref{1} (a). They have the same mass range as the excited states of $c\bar{c}$. The $nnc\bar{n}\bar{n}\bar{c}$ subsystem can be divided into three situations: $(nn)^{I=1}c(\bar{n}\bar{n})^{I=1}\bar{c}$, $(nn)^{ I=0}c(\bar{n}\bar{n})^{I=1}\bar{c}$, and $(nn)^{ I=0}c(\bar{n}\bar{n})^{I=0}\bar{c}$. 

As for $(nn)^{I=1}c(\bar{n}\bar{n})^{I=0}\bar{c}$ states, they are antiparticles of the $(nn)^{ I=0}c(\bar{n}\bar{n})^{I=1}\bar{c}$ states, thus they have the same mass spectra. We find no relative ``stable" states for the $nnc\bar{n}\bar{n}\bar{c}$ system, that is, all of them can decay in $S$-wave through strong interaction.

There are some hexaquark states which have the same quantum numbers among $(nn)^{ I=1}c(\bar{n}\bar{n})^{I=1}\bar{c}$ $(nn)^{I=0}c(\bar{n}\bar{n})^{ I=1}\bar{c}$,  and $(nn)^{ I=0}c(\bar{n}\bar{n})^{ I=0}\bar{c}$. For example, both $(nn)^{ I=1}c(\bar{n}\bar{n})^{ I=1}\bar{c}$ and $(nn)^{I=0}c(\bar{n}\bar{n})^{ I=1}\bar{c}$ have some states with the total isospin ${ I=1}$. The mass spectrum of these states should have been mixed, but all of transition matrix elements of CMI Hamiltonian are zero and thus they cannot be mixed under the CMI model. According to Fig. \ref{1} (a), the masses of $(nn)^{I=1}c(\bar{n}\bar{n})^{ I=1}\bar{c}$ states are usually larger than those of $(nn)^{ I=0}c(\bar{n}\bar{n})^{ I=1}\bar{c}$ states which are generally larger than those of $(nn)^{ I=0}c(\bar{n}\bar{n})^{ I=0}\bar{c}$ states. In the conventional baryon sectors, the $I=1$ one is usually heavier than the $I=0$ one, for example, see [$\Sigma(1189)(I=1)$ vs $\Lambda(1116)(I=0)$] and [$\Sigma_c(2455)(I=1)$ vs $\Lambda_c(2286)(I=0)$].  In our work, the wave functions of hexaquark states can be regarded as “baryon $\otimes$ antibaryon" configuration. These two factors may result into that the hexaquark with larger isospin is heavier than that with smaller isospin. The similar results can be found in Refs. \cite{Chen:2016ont,Weng:2018mmf,Weng:2019ynva}.

The total isospin of $(nn)^{I=1}c(\bar{n}\bar{n})^{I=1}\bar{c}$ states can be $ I=2$, 1, and 0. Note that the symmetry property of $(nn)^{ I=1}c(\bar{n}\bar{n})^{ I=1}\bar{c}$ is determined from $ I_{nn}$ and $I_{\bar{n}\bar{n}}$. Thus, the $(nn)^{ I=1}c(\bar{n}\bar{n})^{ I=1}\bar{c}$ states are degenerate for the total isospin of $ I=2$, $1$, and $0$ in the CMI model.

There are some $nnc\bar{n}\bar{n}\bar{c}$ neutral states with exotic quantum numbers $J^{PC}=0^{--}$, $1^{-+}$, and $3^{-+}$ which the traditional mesons ($q\bar{q}$) cannot have. These exotic quantum number can help identify hidden-charm hexaquark states.

The notation $ H_{n^{2}\bar{n}^{2}}(5036,2^{-},3^{--})$ is for a hexaquark state $nnc\bar{n}\bar{n}\bar{c}$ with mass around 5036 MeV and $I^G(J^{PC})=2^{-}(3^{--})$. According to the Table \ref{eig}, the overlap between $ H_{n^{2}\bar{n}^{2}}(5036,2^{-},3^{--})$ and $\Sigma^{*}_{c}\bar{\Sigma}^{*}_{c}$ states is nearly 1, and thus the hexaquark is mainly made of a baryon and an antibaryon. It may behave like the ordinary scattering state if the inner interaction is not strong, but could also be a resonance or bound state dynamically generated by the baryon and antibaryon. The $ H_{n^2\bar{n}^2}(5060,2^{+},2^{-+})$, $ H_{n^2\bar{n}^2}(5066,2^{+},0^{-+})$, and others have similar situations. These kinds of hexaquarks deserve a more careful study.

The $nnc\bar{n}\bar{n}\bar{c}$ subsystem has one rearrangement decay mode: $nnc-\bar{n}\bar{n}\bar{c}$. The $(nn)^{ I=0}c(\bar{n}\bar{n})^{ I=1}\bar{c}$ hexaquark states can decay to  $\Lambda_{c} \bar{\Sigma}_{c}^{*}$ and $\Lambda_{c} \bar{\Sigma}_{c}$, but the $J^{P}=2^{-}$ ($0^{-}$) states can only decay into $\Lambda_{c} \bar{\Sigma}_{c}^{*}$ ($\Lambda_{c} \bar{\Sigma}_{c}$) due to the angular momentum conservation. The $(nn)^{I=0}c(\bar{n}\bar{n})^{I=0}\bar{c}$ hexaquarks have only one decay channel $\Lambda_{c} \bar{\Lambda}_{c}$ while the $(nn)^{ I=1}c(\bar{n}\bar{n})^{ I=1}\bar{c}$ hexaquark states decay to $\Sigma_{c}^{(*)}\bar{\Sigma}_{c}^{(*)}$ in the ``OZI-superallowed'' decay mode.

One can extract the decay width information from Table \ref{value}.
The $H_{n^2\bar{n}^2}(5143, 2^{-}, 1^{--})$ and $ H_{n^2\bar{n}^2}(5130, 2^{-}, 1^{--})$ decay to all possible channel, but their partial width ratios are different. For the $ H_{n^2\bar{n}^2}(5143, 2^{-}, 1^{--})$,
\begin{equation}	\Gamma_{\Sigma_{c}^{*}\bar{\Sigma}_{c}^{*}}:\Gamma_{(\Sigma_{c}^{*}\bar{\Sigma}_{c})^-}:\Gamma_{\Sigma_{c}\bar{\Sigma}_{c}} = 2.7:1.6:1,
\end{equation}
and for $ H_{n^2\bar{n}^2}(5130, 2^{-}, 1^{--})$·
\begin{equation}	\Gamma_{\Sigma_{c}^{*}\bar{\Sigma}_{c}^{*}}:\Gamma_{(\Sigma_{c}^{*}\bar{\Sigma}_{c})^-}:\Gamma_{\Sigma_{c}\bar{\Sigma}_{c}} = 7.4:11.5:1,
\end{equation}

where $(\Sigma_{c}^{*}\bar{\Sigma}_{c})^-$ is short for the  $(\Sigma_{c}^{*}\bar{\Sigma}_{c} - \Sigma_{c}\bar{\Sigma}_{c}^{*})/\sqrt{2}$ mode with $C=-1$.

\subsection{The $ ssc\bar{s}\bar{s}\bar{c}$ subsystem}
The $ssc\bar{s}\bar{s}\bar{c}$ states can be considered as pure neutral particles. Some of them have normal quantum numbers $J^{PC}=0^{-+}$, $1^{--}$, $2^{-+}$, $2^{--}$, and $3^{--}$, but others carry exotic quantum numbers $J^{PC}=0^{--}$, $1^{-+}$, and $3^{-+}$. Meanwhile, all of these have many different rearrangement decay channels according to Fig. \ref{1} (b) and thus their widths are relative broad.

For the two-body strong decay behaviors of the $ssc\bar{s}\bar{s}\bar{c}$ subsystem, the heaviest $ H_{s^2\bar{s}^2}(5651,0^{+},0^{-+})$ has two decay modes,
\begin{equation}
	\Gamma_{\Omega_{c}^{*}\bar{\Omega}_{c}^{*}}:\Gamma_{\Omega_{c}\bar{\Omega}_{c}}=7.4:1,
\end{equation}
and its dominant decay mode is $\Omega_{c}^{*}\bar{\Omega}_{c}^{*}$. The $H_{s^2\bar{s}^2}(5569,\\0^{-},1^{--})$ state can decay through all possible baryon-antibaryon channels, and
\begin{equation}	\hspace*{-1em}\Gamma_{\Omega_{c}^{*}\bar{\Omega}_{c}^{*}}:\Gamma_{(\Omega_{c}^{*}\bar{\Omega}_{c})^-}:\Gamma_{\Omega_{c}\bar{\Omega}_{c}}=22.6:1.7:1,
\end{equation}
where the $(\Omega_{c}^{*}\bar{\Omega}_{c})^-$ means $(\Omega_{c}^{*}\bar{\Omega}_{c}-\Omega_{c}\bar{\Omega}_{c}^{*})/\sqrt{2}$ which is antisymmetric under the $C$-parity transformation.
Moreover, for the states $ H_{s^2\bar{s}^2}(5429,0^{-},1^{--})$ and $ H_{s^2\bar{s}^2}(5435,0^{+},0^{-+})$, their masses are similar and they can only decay through ${\Omega_{c} \bar{\Omega}_{c}}$ mode. $ H_{s^2\bar{s}^2}(5495,0^{+},0^{-+})$ and $ H_{s^2\bar{s}^2}(5490,0^{+},2^{--})$ have similar masses but the former can decay into ${\Omega_{c} \bar{\Omega}_{c}}$ while the latter can decay through $(\Omega_{c}^{*}\bar{\Omega}_{c}-\Omega_{c}\bar{\Omega}_{c}^{*})/\sqrt{2}$ in $S$-wave.

The rest of $ssc\bar{s}\bar{s}\bar{c}$ hexaquark states are below the baryon-antibaryon decay channels. Therefore, their mainly rearrangement decay channels should be meson-meson-meson decay channels.

\subsection{The $nnc\bar{s}\bar{s}\bar{c}$ subsystem}
According to Fig. \ref{1} (c), we discuss the mass spectra and decay behaviour of $nnc\bar{s}\bar{s}\bar{c}$ subsystem.
For the $I=1$ $nnc\bar{s}\bar{s}\bar{c}$ states, they are explicitly exotic states. There are still no relative stable states in $nnc\bar{s}\bar{s}\bar{c}$ subsystem.

For the $ I=0$ states, they have only two channels: ${\Lambda_c\bar{\Omega}_c^{*}}$ and ${\Lambda_c\bar{\Omega}_c}$. The two states $ H_{n^2\bar{s}^2}(5128,0,2^{-})$ and $H_{n^2\bar{s}^2}(5130,0,0^{-})$ can be distinguished by their respective decay modes. The $H_{n^2\bar{s}^2}(5128,0,2^{-})$ can dissociate into $\Lambda_c \bar{\Omega}_c^* $ while $H_{n^2\bar{s}^2}(5130,0,0^{-})$ can decay into $\Lambda_c \bar{\Omega}_c $ in $S$-wave.

There are four different decay channels for the $I=1$ states: ${\Sigma_c^{*}\bar{\Omega}_c^{*}}$, ${\Sigma_c\bar{\Omega}_c^{*}}$, ${\Sigma_c^{*}\bar{\Omega}_c}$, and ${\Sigma_c \bar{\Omega}_c}$.  From Table \ref{value}, for $H_{n^2\bar{s}^2}(5415,1,1^{-})$ state,
\begin{equation}
	\Gamma_{\Sigma_c^* \bar{\Omega}_c^{*}}:\Gamma_{\Sigma_c \bar{\Omega}_c^{*}}:\Gamma_{\Sigma_c^* \bar{\Omega}_c}:\Gamma_{\Sigma_c \bar{\Omega}_c}=9.4:2.2:1.9:1.
\end{equation}
and for the $H_{n^2\bar{s}^2}(5312,1,1^{-})$ state 
\begin{equation}
	\Gamma_{\Sigma_c^* \bar{\Omega}_c^{*}}:\Gamma_{\Sigma_c \bar{\Omega}_c^{*}}:\Gamma_{\Sigma_c^* \bar{\Omega}_c}:\Gamma_{\Sigma_c \bar{\Omega}_c}=1.6:0.2:8:1.
\end{equation}
%

\subsection{The $nsc\bar{n}\bar{n}\bar{c}$ subsystem}

We discuss the mass spectra and decay behaviors of $nsc\bar{n}\bar{n}\bar{c}$ subsystem based on Fig. \ref{2} (a). The $nnc\bar{n}\bar{s}\bar{c}$ states are antiparticles of the $nsc\bar{n}\bar{n}\bar{c}$ states, and thus they have the same mass spectra.

The $nsc\bar{n}\bar{n}\bar{c}$ subsystem can be divided into two situations: $nsc(\bar{n}\bar{n})^{ I=1}\bar{c}$ and $ nsc(\bar{n}\bar{n})^{ I=0}\bar{c}$. For the  $nsc(\bar{n}\bar{n})^{ I=1}\bar{c}$ states, the mass spectra are identical for total isospin of $I=3/2$ and $1/2$ in CMI model similar to $(nn)^{ I=1}c(\bar{n}\bar{n})^{ I=1}\bar{c}$ subsystem. The $nsc\bar{n}\bar{n}\bar{c}$ states with $I=3/2$ are explicitly exotic and thus easily identifiable as candidates for the hidden-charm hexaquark state.

From Fig. \ref{2} (a), we find the lowest $0^{-}$, $1^{-}$, $2^{-}$, and $3^{-}$ states are relatively stable states, and especially the $ H_{ns\bar{n}^{2}}(3578, 1/2, 0^{-})$ is below all the thresholds for rearrangement decay channels. Other three states can still decay via $D$-wave strong interaction. For example, the $ H_{ns\bar{n}^{2}}(4682,{1}/{2},3^{-})$ can decay into $KD^{*}\bar{D}$ final states via $D$-wave.

From Table \ref{value}, there are 6 and 3 possible baryon-antibaryon channels for the $nsc(\bar{n}\bar{n})^{ I=1}\bar{c}$ and $nsc(\bar{n}\bar{n})^{ I=0}\bar{c}$ subsystems, respectively. The $H_{ns\bar{n}^2}(5071,3/2,2^{-})$ and $H_{ns\bar{n}^{2}}(5071,3/2,0^{-})$ are accidentally degenerate, but the $J^P=2^{-}$ state can decay into $\Xi_{c}^* \bar{\Sigma}_{c}^*$, $\Xi_{c}^* \bar{\Sigma}_{c}$, $\Xi_{c}' \bar{\Sigma}_{c}^*$, and $\Xi_{c}' \bar{\Sigma}_{c}$ while the $J^P=0^{-}$ state can only decay through  $\Xi_{c} \bar{\Sigma}_{c}^*$, and $\Xi_{c} \bar{\Sigma}_{c}$ channels. Similarly, $H_{ns\bar{n}^2}(4971,1/2,2^{-})$ can only decay through ${\Xi_{c}^{*} \bar{\Lambda}_{c}}$ mode, but $H_{ns\bar{n}^2}(4970,1/2,1^{-})$ can only decay into ${\Xi_{c}' \bar{\Lambda}_{c}}$, ${\Xi_{c}' \bar{\Lambda} _{c}}$
and ${\Xi_{c} \bar{\Lambda}_{c}}$ modes. We can distinguish $H_{ns\bar{n}^2}(4913,1/2,1^{-})$ and $H_{ns\bar{n}^2}(4914,1/2,0^{-})$ by partial decay width ratios since for the former,
\begin{equation}
	\Gamma_{\Xi_{c}' \bar{\Lambda} _{c}}:\Gamma_{\Xi _{c} \bar{\Lambda} _{c}} = 0.1:1.
\end{equation}
and for the latter,
\begin{equation}
	\Gamma_{\Xi_{c}' \bar{\Lambda} _{c}}:\Gamma_{\Xi _{c} \bar{\Lambda} _{c}} = 4.6:1.
\end{equation}
%

\subsection{The $nsc\bar{n}\bar{s}\bar{c}$ subsystem}
Here, we discuss the $nsc\bar{n}\bar{s}\bar{c}$ subsystem based on Fig. \ref{2} (b). The subsystem is also a pure neutral subsystem, thus $C$ parity and $G$ parity are good quantum numbers. Since there is no constraint from the Pauli principle for $nsc\bar{n}\bar{s}\bar{c}$ subsystem, the values of $\delta^{A}_{12}$, $\delta^{S}_{12}$, $\delta^{A}_{34}$, and $\delta^{S}_{34}$ from Table \ref{type} are all 1 and the obtained mass spectra is more complicated than other subsystems. There are no relative stable states for the $nsc\bar{n}\bar{s}\bar{c}$ subsystem.

Similar to the $nnc\bar{n}\bar{n}\bar{c}$ states, the mass spectra of $nsc\bar{n}\bar{s}\bar{c}$ states are identical for total isospin of $I=1$ and $0$ in CMI model. From Fig. \ref{2} (b), we find nine good exotic states candidates for quantum numbers $J^{PC}=0^{--}$.

The mass of $H_{ns\bar{n}\bar{s}}(5117,1^{-}(0^{+}),0^{--})$ is very closed to $H_{ns\bar{n}\bar{s}}(5114,1^{-}(0^{+}),1^{--})$, and they all can decay into $(\Xi_{c}'\bar{\Xi}_{c}-\Xi_{c}\bar{\Xi}_{c}')/\sqrt{2}$. However,  the $H_{ns\bar{n}\bar{s}}(5114,1^{+}(0^{-}),1^{--})$ has others decay channels. From Table \ref{value}, we obtain for $H_{ns\bar{n}\bar{s}}(5114,1^{+}(0^{-}),1^{--})$
\begin{eqnarray}
&&\Gamma_ {(\Xi _{c}^{*} \bar{\Xi} _{c})^{-}}:\Gamma_{(\Xi _{c}'\bar{\Xi} _{c})^{-}}:\Gamma_{\Xi_{c} \bar{\Xi}_{c}}=0.6:18.1:1,
\end{eqnarray}

where $(\Xi _{c}^{*} \bar{\Xi} _{c})^{-}$ and $(\Xi _{c}'\bar{\Xi} _{c})^{-}$ represent $(\Xi _{c}^{*} \bar{\Xi} _{c}-\Xi_{c}\bar{\Xi}_{c}^{*})/\sqrt{2}$ and $(\Xi_{c}'\bar{\Xi}_{c}-\Xi_{c}\bar{\Xi}_{c}')/\sqrt{2}$ respectively.

\subsection{The $nsc\bar{s}\bar{s}\bar{c}$ subsystem}

Lastly, we discuss the mass spectra and decay behaviour of $nsc\bar{s}\bar{s}\bar{c}$ subsystem based on the Fig. \ref{3} (a). The $ssc\bar{n}\bar{s}\bar{c}$ states are antiparticles of the $nsc\bar{s}\bar{s}\bar{c}$ states, and thus they have the same mass spectra. The restrictions from Pauli principle for the $nsc\bar{s}\bar{s}\bar{c}$ states are the same as the $nsc(\bar{n}\bar{n})^{I=1}\bar{c}$ states, and therefore the numbers of their states are equal.

From Fig. \ref{3} (a), we easily find that there are no relative stable states. The $nsc\bar{s}\bar{s}\bar{c}$ states are higher than many different rearrangement decay channels. Therefore, they would have a relative wide width. In conclusion, we do not suggest that the experimentalists foremost find these states.

The reference baryon-antibaryon systems for the $nsc\bar{s}\bar{s}\bar{c}$ states are the $\Xi^{*}_{c}\bar{\Omega}^{*}_{c}$, $\Xi'_{c}\bar{\Omega}^{*}_{c}$, $\Xi_{c}\bar{\Omega}^{*}_{c}$, $\Xi^{*}_{c}\bar{\Omega}_{c}$, $\Xi'_{c}\bar{\Omega}_{c}$, and $\Xi_{c}\bar{\Omega}_{c}$. The mass of $H_{ns\bar{s}^2}(5356,1/2,2^-)$ is close to that of $H_{ns\bar{s}^2}(5354,1/2,0^-)$. For $H_{ns\bar{s}^2}(5356,1/2,2^-)$ state, it can decay through ${\Xi_{c}'\bar{\Omega}_{c}^{*}}$, ${\Xi_{c}^{*}\bar{\Omega}_{c}}$, and ${\Xi_{c}\bar{\Omega}_{c}}$ channels. But $H_{ns\bar{s}^2}(5354,1/2,0^-)$ can decay into ${\Xi_{c}'\bar{\Omega}_{c}}$ and ${\Xi_{c}\bar{\Omega}_{c}}$ channels. Then we consider $H_{ns\bar{s}^2}(5245,1/2,2^-)$ and $H_{ns\bar{s}^2}(5244,1/2,2^-)$, and both of them can decay through ${\Xi_{c}\bar{\Omega}_{c}^{*}}$ and ${\Xi_{c}\bar{\Omega}_{c}}$ channels in $S$-wave. From Table \ref{value}, for $H_{ns\bar{s}^2}(5245,1/2,2^-)$,
\begin{eqnarray}
\Gamma_{\Xi_{c}\bar{\Omega}_{c}^{*}}:\Gamma_{\Xi_{c}\bar{\Omega}_{c}}=0.1:1,
\end{eqnarray}
and for $H_{ns\bar{s}^2}(5244,1/2,2^-)$,
\begin{eqnarray}
\Gamma_{\Xi_{c}\bar{\Omega}_{c}^{*}}:\Gamma_{\Xi_{c}\bar{\Omega}_{c}}=0.3:1.
\end{eqnarray}

\section{SUMMARY}
\label{sec5}

Up to now, more and more hidden-charm tetraquark states and pentaquark states have been discovered and confirmed by different experiments.
These give us a significant confidence to the existence of hidden-charm hexaquark states.
Thus, we studied systemically the mass spectra, stability, and strong decay behaviors of hidden-charm hexaquark states in the framework of the CMI model.

\begin{table}[htbp]
	\caption{The relatively stable states of hidden-charm hexaquark system in CMI model. The masses are all in units of MeV.}\label{aaa}
	\renewcommand\arraystretch{1.55}
	\renewcommand\tabcolsep{3.75pt}
	\begin{tabular}{|c|cc|c|cc|}
		\bottomrule[1.5pt]
		\bottomrule[0.5pt]
		States&$I^{G}(J^{PC})$&Masses&States&$I^{(G)}(J^{P(C)})$&Masses\\
		\bottomrule[0.7pt]
		
		\multirow{6}*{$nsc\bar{n}\bar{s}\bar{c}$}&$0^{+}(0^{-+})$&3815& \multirow{4}*{$nsc\bar{n}\bar{n}\bar{c}$}& $1/2(0^{-})$& 3578\\ \cline{2-3} \cline{5-6}
		&$0^{-}(1^{--})$&4005&&$1/2(1^{-})$&3670\\ \cline{2-3}\cline{5-6}
		&$0^{-}(2^{--})$&\multirow{2}*{4443}&&$1/2(2^{-})$&4090\\  \cline{5-6}
		&$1^{+}(2^{--})$&&&$1/2(3^{-})$&4682\\ \cline{2-6}
		&$0^{-}(3^{--})$&\multirow{2}*{4794}&\multirow{2}*{$nnc\bar{n}\bar{n}\bar{c}$}&{$0^{+}(1^{-+})$}&{3887} \\
		&$1^{+}(3^{--})$& &\multicolumn{1}{c|}{} & \multicolumn{1}{c}{$0^{-}(3^{--})$}&4576 \\
		\cline{2-3} \cline{1-6}
		\bottomrule[0.5pt]
		\midrule[1.5pt]
	\end{tabular}
\end{table}

Firstly, we introduce the CMI model and extract the corresponding coupling constants from traditional hadrons.
Next, we construct the flavor $\otimes$ color $\otimes$ spin wavefunctions based on the SU(3) and SU(2) symmetry.
Meanwhile, we require the wavefunction to obey Pauli Principle.
After that, we systemically calculate the mass spectra, corresponding overlap, and the values of $k\cdot|c_{i}|$. Lastly, we specifically discuss the stability, the possible quark rearrangement decay channels, and the relative decay width ratios.

For $nnc\bar{n}\bar{n}\bar{c}$, $ssc\bar{s}\bar{s}\bar{c}$, and $nsc\bar{n}\bar{s}\bar{c}$ subsystems, they are pure neutral particles (except $(nn)^{ I=0}c(\bar{n}\bar{n})^{ I=1}\bar{c}$ subsystem), and $C$ parity and $G$ parity both are good quantum numbers.
According to the mass spectra, we find that the lower isospin quantum number, the more compact hexaquark states. Here, the $J^{PC}=0^{--}, 1^{-+}, 3^{-+}$ states are good exotic states candidates, and especially the $0^{--}$ states which even the $S$-wave tetraquark states cannot carry.

We list some possible stable hexaquark states in Table \ref{aaa}. We find ten relative stable states, which are below all allowed rearrangement decay channels. These states belong to the $nnc\bar{n}\bar{n}\bar{c}$ subsystem, $nsc\bar{n}\bar{n}\bar{c}$ subsystem and  $nsc\bar{n}\bar{s}\bar{c}$ subsystem respectively. We think the $H_{ns{\bar{n}}^{2}}(3578, 1/2, 0^{-})$ and $H_{ns{\bar{n}}^{2}}(3670, 1/2, 1^{-})$ states are better stable candidates which could be first searched for in experiments.

\begin{table}[htbp]
	\centering
	\caption{The comparison of the masses for the $(nn)^{I=0}c(\bar n\bar n)^{I=0}\bar c$ and $ssc \bar s\bar s\bar c$ systems with $I^G(J^{PC})=0^-(1^{--})$ in two scenarios. All masses are in units of MeV.}
	\label{uncertainty}
	\renewcommand\arraystretch{1.3}
	\renewcommand\tabcolsep{15.5pt}
	\begin{tabular}{cc|cc}
		\bottomrule[1.5pt]
		\bottomrule[0.5pt]
		$nnc \bar n\bar n\bar c$   & & $ssc \bar s\bar s\bar c$   & \\
		\bottomrule[0.7pt]
		Scen.1 & Scen.2 & Scen.1 & Scen.2 \\
		\bottomrule[0.7pt]
		3600  & 3201  & 4836  & 4670   \\
		3736  & 3443  & 4940  & 4871   \\
		4197  & 3960  & 5100  & 5068   \\
		4234  & 4053  & 5175  & 5152   \\
		4325  & 4260  & 5275  & 5258   \\
		4432  & 4437  & 5287  & 5274   \\
		4548  & 4520  & 5329  & 5298   \\
		4584  & 4588  & 5429  & 5434   \\
		4816  & 4835  & 5522  & 5556   \\
		4940  & 5003  & 5569  & 5592   \\
		\bottomrule[0.5pt]
		\bottomrule[1.5pt]
	\end{tabular}
\end{table}

In order to check the uncertainty of our framework, we also determine 
the $v_{q\bar{q}}$ and $m_{q\bar{q}}$ with the masses of pseudoscalar 
mesons. Since the spontaneously breaking of vacuum symmetry strongly 
affects the properties of these pseudoscalar mesons, the parameters of 
$v_{q\bar{q}}$ and $m_{q\bar{q}}$ are not the same as those obtained 
with the vector mesons. For example, $v_{n\bar{n}}$ and $m_{n\bar{n}}$ 
become 29.87 MeV and 153.99 MeV in the new scenario, respectively. However, the 
difference between the hexquark masses of the two scenarios can be 
roughly used to estimate the uncertainly of our approach. We give the 
comparison of the $(nn)^{I=0}c(\bar n\bar n)^{I=0}\bar c$ and $ssc \bar s\bar s\bar c$ systems with $I^G(J^{PC})=0^-(1^{--})$ in 
Table \ref{uncertainty}. Scen.1 (Scen.2) denotes the results calculated by using the parameters obtained with the vector (pseudoscalar) mesons. Firstly, one notices that the ground states differ largest from the table. The heavier the state is, the smaller the difference between the two scenarios is. These may be resulted from that the new $v_{q\bar{q}}$ becomes larger while the new $m_{q\bar{q}}$ becomes smaller. Secondly, the mass of the $(nn)^{I=0}c(\bar n\bar n)^{I=0}\bar c$ ground state with $I^G(J^{PC})=0^-(1^{--})$ changes about 399 MeV while that for the $ssc \bar s\bar s\bar c$ case only varies about 166 MeV. That is, the uncertainty reduces when the number of $n/\bar{n}$ quark in hexaquark states decreases, which is because the mass difference between $\pi$ and $\rho$ mesons is much larger than those between $K$ and $K^*$ mesons.

In summary, we give a preliminary study about the mass spectra of hidden-charm hexaquark states. In addition to the CMI model, other non-perturbative QCD methods can also help us to understand more properties of the hexaquark states in detail such as QCD sum rule, effective fields theories and lattice QCD simulations. We hope that our study may inspire theorists and experimentalists to pay attention to these hidden-charm hexaquark states.

\section{Acknowledgments}
This work is supported by the China National Funds for Distinguished Young Scientists under Grant No. 11825503, National Key Research and Development Program of China under Contract No. 2020YFA0406400, the 111 Project under Grant No. B20063, and the National Natural Science Foundation of China under Grant No. 12047501. This project is also supported by the National Natural Science Foundation of China under Grants No. 12175091, and 11965016, CAS Interdisciplinary Innovation Team, and the Fundamental Research Funds for the Central Universities under Grants No. lzujbky-2021-sp24.


\end{document}